\documentclass[twocolumn]{aastex631}

\usepackage{hyperref}
\usepackage{amsmath}

\begin{document}

\title{Interior and Gravity Field Models for Uranus Suggest Mixed-composition Interior:\\ Implications for the Uranus Orbiter and Probe}

\shorttitle{Interior and Gravity Models for Uranus}


\correspondingauthor{Zifan Lin}
\email{zifanlin@mit.edu}

\author[0000-0003-0525-9647]{Zifan Lin}
\affiliation{Department of Earth, Atmospheric, and Planetary Sciences, Massachusetts Institute of Technology, 77 Massachusetts Avenue, Cambridge, MA 02139, USA}

\author[0000-0002-6892-6948]{Sara Seager}
\affiliation{Department of Earth, Atmospheric, and Planetary Sciences, Massachusetts Institute of Technology, 77 Massachusetts Avenue, Cambridge, MA 02139, USA}
\affiliation{Department of Physics and Kavli Institute for Astrophysics and Space Research, Massachusetts Institute of Technology, Cambridge, MA 02139, USA}
\affiliation{Department of Aeronautics and Astronautics, MIT, 77 Massachusetts Avenue, Cambridge, MA 02139, USA}

\author[0000-0003-3113-3415]{Benjamin P. Weiss}
\affiliation{Department of Earth, Atmospheric, and Planetary Sciences, Massachusetts Institute of Technology, 77 Massachusetts Avenue, Cambridge, MA 02139, USA}





\begin{abstract}
The interior composition and structure of Uranus are ambiguous. It is unclear whether Uranus is composed of fully differentiated layers dominated by an icy mantle or has smooth compositional gradients. The Uranus Orbiter and Probe (UOP), the next NASA Flagship mission prioritized by the Planetary Science and Astrobiology Survey 2023--2032, will constrain the planet's interior by measuring its gravity and magnetic fields. To characterize the Uranian interior, here we present \texttt{CORGI}, a newly developed planetary interior and gravity model. We confirm that high degrees of mixing are required for Uranus interior models to be consistent with the $J_2$ and $J_4$ gravity harmonics measured by Voyager 2. Empirical models, which have smooth density profiles that require extensive mixing, can reproduce the Voyager 2 measurements. Distinct-layer models with mantles composed of H$_2$O-H/He or H$_2$O-CH$_4$-NH$_3$ mixtures are consistent with the Voyager 2 measurements if the heavy element mass fraction, $Z$, in the mantle $\lesssim85\%$, or if atmospheric $Z$ $\gtrsim25\%$. Our gravity harmonics model shows that UOP $J_2$ and $J_4$ measurements can distinguish between high ($Z\geq25\%$) and low ($Z=12.5\%$) atmospheric metallicity scenarios. The UOP can robustly constrain $J_6$ and potentially $J_8$ given polar orbits within rings. An ice-rich composition can naturally explain the source of Uranus' magnetic field. However, because the physical properties of rock-ice mixtures are poorly known, magnetic field generation by a rock-rich composition cannot be ruled out. Future experiments and simulations on realistic planetary building materials will be essential for refining Uranus interior models.
\end{abstract}





\section{Introduction} \label{sec:intro}

Uranus and Neptune represent a unique intermediate-sized end-member population of planetary evolution, but their structures and compositions are ambiguous. In general, there are two classes of interior models for Uranus and Neptune: distinct-layer structures and empirical density profiles.

Distinct-layer structures are the first and the most common class of models. Typically, such models include three fully-differentiated, adiabatic layers: a small rocky core ($\lesssim20\%$ by radius), a thick icy mantle making up the majority of the planet's mass and volume ($\gtrsim70$\% by mass and $\sim50\%$ by radius), and an atmosphere dominated by hydrogen and helium (H/He) with a thickness of about 30\% of the planetary radius \citep[e.g.,][]{fortney_interior_2010, nettelmann_new_2013}. Distinct-layer models have the advantage of being self-consistent and physically motivated: the material composition, pressure, temperature, and density at a certain depth in a planet model can always be uniquely defined by employing a physical equation of state (EOS) and an adiabatic pressure-temperature (P-T) profile. However, distinct-layer models by definition fail to capture possible mixing between layers. Furthermore, it is biased towards an ice-rich composition due to the ice-like bulk densities of Uranus and Neptune: a rock-rich composition is missed by distinct-layer models \citep[see e.g.,][for the debate between ice-rich and rock-rich interior compositions for Uranus and Neptune]{helled_interiors_2020, teanby_neptune_2020, vazan_new_2022}.

The second class of models create so-called empirical density profiles. This class of models is motivated by planet formation theories that suggest the deep interior of Uranus and Neptune should contain compositional gradients \citep[e.g.,][]{helled_interiors_2020, vazan_explaining_2020}. Instead of solving for physical equations, empirical density models generate a wide range of monotonic functions of density, $\rho(r)$, that match the observed mass, radius, and gravity harmonics of Uranus \citep[e.g.,][]{marley_monte_1995, podolak_further_2000, helled_interior_2011, podolak_random_2022, movshovitz_promise_2022, neuenschwander_empirical_2022, morf_interior_2024}. Physical interpretations are not offered for these density profiles -- the density is estimated while being agnostic to the material composition and thermal state at a certain depth. Empirical density models are nevertheless helpful for probing parameter spaces that are missed by distinct-layer models, especially complex interiors with compositional gradients and non-adiabatic temperature profiles. Unifying the distinct-layer and empirical density approaches requires insights into the unknown thermodynamic behaviors of mixtures.

The interior structure degeneracies of Uranus and Neptune are exemplary of a general compositional degeneracy among intermediate-sized exoplanets in the Galaxy.
While both the smallest, densest planets (Earth-like planets and super-Earths dominated by iron and silicates) and the largest, least dense planets (gas giants dominated by H/He) have well-defined interior compositions, planets with intermediate sizes and densities are much more poorly understood.
The masses and radii of intermediate-sized planets are compatible with a wide range of internal structures, including super-Earths with thick H/He envelopes \citep[e.g.,][]{valencia_internal_2006, seager_massradius_2007, wagner_interior_2011, zeng_detailed_2013, boujibar_superearth_2020}, Uranus- and Neptune-like planets with volatile-ice-rich interiors and substantial H/He envelopes \citep[e.g.,][]{rogers_framework_2010, nettelmann_thermal_2011, valencia_bulk_2013}, and water worlds almost completely dominated by H$_2$O with little or no H/He \citep[e.g.,][]{sotin_massradius_2007, thomas_hot_2016, madhusudhan_interior_2020, luque_density_2022, rigby_ocean_2024}. 

\subsection{Past and Future Spacecraft Measurements}
Future spacecraft measurements of Uranus' gravity and magnetic fields offer the opportunity of resolving the planet's interior composition degeneracy. 
In a single flyby in 1986, Voyager 2 (V2) measured the $J_2$ and $J_4$ gravity harmonics of Uranus.
These measurements significantly constrain the range of possible mass distributions within Uranus compared to knowing only the mass and radius \citep{movshovitz_promise_2022}. 
Prioritized as the next Flagship mission by the Planetary Science and Astrobiology Decadal Survey 2023--2032 \citep[][hereafter the decadal survey]{decadal_survey_2023}, the Uranus Orbiter and Probe (UOP) mission offers a rare opportunity to precisely measure the gravity field of Uranus.

Precise gravity measurements have previously placed powerful constraints on the interior structures of Jupiter and Saturn. Before the Juno and Cassini gravity field measurements, it was long debated whether the heavy element cores of gas giants are compact or dilute (see \citealt{stevenson_jupiters_2020} for a review). Precise gravity field measurements from Cassini's Grand Finale constrained its core mass, $\sim$15--18 $M_\oplus$, and heavy element mass distributed throughout the envelope, 1.5--5 $M_\oplus$ \citep{militzer_models_2019}. Jupiter's gravity field measured by Juno up to $J_{12}$ \citep{iess_measurement_2018} implies a dilute core \citep{militzer_juno_2022, howard_jupiters_2023}. In addition, precise gravity measurements helped constrain deep atmosphere dynamics in the gas giants. Cassini measured unexpectedly large values of $J_6$, $J_8$, and $J_{10}$ (approximately 7\%, 1.6 times, and 4.3 times larger than that expected for a uniformly rotating interior respectively), implying differential rotation of the deep atmosphere \citep{iess_measurement_2019, militzer_models_2019}. Likewise, precise gravity measurements by UOP can potentially resolve the controversy around the interior composition of Uranus or reveal its previously unknown deep atmospheric motion in a similar manner.

Further, V2 acquired measurements of Uranus' intrinsic magnetic field, revealing a surprising multipolar, non-axisymmetric geometry \citep{ness_magnetic_1986, ness_magnetic_1989}. This discovery suggested that Uranus' magnetic field may be generated in a shallow convective thin shell \citep{stanley_convective-region_2004, stanley_numerical_2006}. Uranus interior models that include a layer of conducting fluid in convective motion at depths predicted by the convective thin shell geometry may therefore be preferred.

Future magnetic field measurements by the UOP can reveal the depth and thickness of dynamo generating region inside Uranus, providing indirect evidence for its interior structure and composition. Similar to the gravity field, magnetic field of a planet can be decomposed into spherical harmonics that decays as $1/r^{n+1}$, where $n$ is the degree and $n=1$ represents the dipole. Due to this rapid $1/r^{n+1}$ decay, the multipolar magnetic field of Uranus likely originates from a shallow region. Otherwise, the magnetic field observed by V2 would likely to be dominated by the dipolar component. The multipolar field of Uranus and Neptune was the motivation for developing the convective thin shell geometry \citep{stanley_convective-region_2004, stanley_numerical_2006}. \cite{soderlund_underexplored_2020} explored a wider parameter space assuming similar structure by varying the core size and convective thin shell size. UOP magnetic field measurements will place tighter constraints on dynamo model parameters, in turn narrowing down the parameter space of allowed interior structure and composition models. 

Here we present \texttt{CORGI} (Composition Of Rocky, Gaseous, and Icy planets), a code package with three modules: a forward planet interior structure module capable of generating both distinct-layer and empirical density planet models, an inverse module for retrieving the possible compositions of a planet given its mass and radius, and a gravity harmonics module adopting the concentric Maclaurin spheroid (CMS) method \citep{hubbard_concentric_2013}. We start by generating a variety of Uranus interior models permitted by its mass and radius using both the distinct-layer and empirical density approaches. Then, we simulate the high-precision gravity harmonics for all these interior models and discuss the implications for the UOP mission. Section \ref{sec:methods} introduces underlying physics of the \texttt{CORGI} code package. We present our major findings in Section \ref{sec:results} and discuss implications of our results in Section \ref{sec:discussion}. Our conclusions are summarized in Section \ref{sec:conclusion}.


\section{Methods} \label{sec:methods}
We now introduce the three modules of \texttt{CORGI}. In Section \ref{sec:interior_module}, we describe the distinct-layer planetary interior forward model. In Section \ref{sec:empirical_density_model}, we describe the empirical density forward model. In Section \ref{sec:inverse_model}, we summarize the inverse model that retrieves the most probable interior composition of planet given exterior constraints. Section \ref{sec:CMS_model_details} outlines the CMS model for simulating zonal gravity harmonics of a uniformly rotating planet.

\subsection{Distinct-layer Forward Model} \label{sec:interior_module}

Here we describe the distinct-layer interior structure module of \texttt{CORGI}. Section \ref{sec:distinct_layer_model} introduces the underlying physics of the distinct-layer model. Section \ref{sec:eos} presents EOSs of materials incorporated in the distinct-layer model and introduces how we calculate EOSs of mixtures. We then discuss how the interior temperature profiles of distinct-layer planets are modelled in Section \ref{sec:temp_profile}. Finally, we validate the model in Section \ref{sec:distinct_layer_validation}. 

\subsubsection{Model Setup} \label{sec:distinct_layer_model}
The interior structure of a nonrotating, spherically symmetric planet can be solved by three fundamental equations \citep{zapolsky_mass-radius_1969}, namely the mass of a spherical shell
\begin{equation} \label{eq:mass_in_shell}
    \frac{dm(r)}{dr} = 4 \pi r^2 \rho(r),
\end{equation}
hydrostatic equilibrium
\begin{equation} \label{eq:hydro_eq}
    \frac{dP(r)}{dr} = -\frac{Gm(r)\rho(r)}{r^2},
\end{equation}
and EOS
\begin{equation} \label{eq:eos}
    P(r) = f(\rho(r), T(r)).
\end{equation}
Equation (\ref{eq:eos}) relies on a temperature function that will be discussed in Section \ref{sec:temp_profile}. We start from the core of the planet assuming some central pressure, $P_{\rm c}$, and numerically integrate outwards with a default step size of 100 m until reaching the desired planetary mass, $M_p$, and radius, $R_p$. 
To avoid a vanishing mass at $r=0$, the core is treated as a small constant-density sphere with a default radius, $r_{\rm c}$, of $10$ m.

By default, \texttt{CORGI} assumes a four-layer planet with a Fe core, a MgSiO$_3$ mantle, an overlying H$_2$O layer, and an H/He envelope. The user may opt to remove one or more layers (e.g., to model a terrestrial planet with only iron and silicate layers). Additional layers can be added with user-supplied EOSs. For a specified composition $\{x_i\}$ (e.g., $\{x_{\rm Fe}, x_{\rm MgSiO_3}, x_{\rm mantle}, x_{\rm atm}\}$ for the default four-layer planet), where $x_i$ denotes mass fraction of component $i$ and $\sum x_i = 1$, the code iterates until a layer's mass reaches $x_i M_p$ before switching to the next layer.

The outer boundary condition is simply $M(r) = M_p$ if the outermost layer is not an H/He envelope. Note that for a given composition, $\{x_i\}$, $r$ does not necessarily equal $R_p$ when $M_p$ is reached, if the initial $P_c$ guess is inaccurate. We employ a shooting method to solve the boundary value problem: we run the forward model iteratively until the right $P_{\rm c}$ that makes $M(R_p) = M_p$ is found by bisection.

If the outermost layer is an H/He envelope, then because the surface of a gaseous envelope is not well defined, we add an extra outer boundary condition that the optical depth $\tau_t(R_p) = 1$, where the subscript $t$ denotes the transverse optical depth through the limb of the planet \citep[following][]{rogers_framework_2010}. This translates into an exterior boundary condition for the radial optical depth at $R_p$, denoted as $\tau_R$, as
\begin{equation}
    \tau_R = \frac{1}{\gamma} \sqrt{\frac{H_p}{2\pi (\alpha+1) R_p}},
\end{equation}
where $\gamma \equiv \kappa_{\rm v} / \kappa_{\rm th}$ is the ratio between the optical and infrared optical depths and $H_p$ is the constant pressure scale height defined as
\begin{equation}
    H_p = \frac{R_p^2 k_B T}{G M_p \mu_{\rm eff}},
\end{equation}
where $\mu_{\rm eff}$ is the effective molecular mass of the gas. 
Given the optical depth at $R_p$, the pressure at $R_p$ is expressed as
\begin{equation}
    P_R = \left[ \frac{G M_p (\alpha+1) \tau_R}{R_p^2 C T^\beta} \right]^{1/(\alpha+1)}.
\end{equation}
In the above expression, $\log C = -7.32$, $\alpha = 0.68$, and $\beta = 0.45$ are derived from fitting tabulated Rosseland mean opacities for H/He from \cite{freedman_line_2008}. 

The user may also choose to match a temperature outer boundary condition, such as $T(R_p) = T_{\rm eq}$. This is achievable by tuning the central temperature $T_{\rm c}$ using the same shooting method used to find the central pressure (see details for temperature profile in Section \ref{sec:temp_profile}).

\subsubsection{Equations of State (EOS)} \label{sec:eos}
Here we discuss the default EOSs employed in \texttt{CORGI} for each material. Following previous studies, for Fe and MgSiO$_3$, \texttt{CORGI} adopts isothermal EOSs because the densities of iron and rocks under high pressures are not sensitive to temperature changes. The user can choose the Vinet EOS \citep{vinet_compressibility_1987, vinet_universal_1989}, Birch-Murnagham EOS \citep{birch_finite_1947, poirier_introduction_2000}, which are both fit to experimental data, or the adapted polynomial EOS \citep{holzapfel_equations_1998, holzapfel_coherent_2018}. Furthermore, for terrestrial planets at low pressures, \texttt{CORGI} also offers the option of the preliminary reference Earth model (PREM) \citep{zeng_massradius_2016}. At the high pressure limit ($\gtrsim 10^4$ GPa), all EOSs converge to the quantum mechanical Thomas-Fermi-Dirac theory \citep{salpeter_theoretical_1967}.  All the above EOSs match experimental data reasonably well. Differing choices of the EOS will change the predicted radius of a rocky planet by no more than 1.5\% at $\sim 10\,M_\oplus$, smaller than the typical exoplanet radius error ($<8\%$ is considered ``precise," see e.g., \citealt{luque_density_2022}).  \texttt{CORGI} provides all the above EOSs, but its default option is the adapted polynomial EOS for Fe and MgSiO$_3$ \citep{zeng_new_2021}. 

The presence of H$_2$O introduces large uncertainties into interior structure models due its numerous phase transitions and the sensitive temperature-dependence of its density. The supercritical phase of water occupies a large portion of the P-T parameter space and expands significantly with temperature, leading to major changes in the predicted planetary mass and radius \citep[e.g.,][]{mousis_irradiated_2020, nixon_how_2021}. High-pressure water ice becomes superionic at pressures and temperatures relevant for the interiors of Uranus and Neptune \citep{millot_nanosecond_2019}, which is likely important for their dynamo generation mechanism. To capture the effects of this sensitive pressure and temperature dependence, the default water EOS in \texttt{CORGI} is AQUA, which uses thermodynamically consistent interpolation to a span a wide P-T-range (0.1 Pa to 400 TPa and 150 to $10^5$ K) through incorporation of published H$_2$O EOSs developed for more limited P-T conditions \citep{haldemann_aqua_2020}. 
For H/He, \texttt{CORGI} adopts a recent wide P-T-range EOS database from $10^{-9}$ to $10^{13}$ GPa for pressures and from $10^2$ to $10^8$ K for temperatures \citep{chabrier_new_2019, chabrier_new_2021}. By default, a solar helium mass fraction of $0.275$ is assumed.

In the interiors of realistic planets, the ice layer is unlikely to be pure H$_2$O and the atmosphere is unlikely to be pure H/He. Previous models of Uranus and Neptune generally assume some fraction of light elements mixing into the ice layer and some fraction of heavy elements mixing into the atmosphere \citep[e.g.,][]{fortney_interior_2010, nettelmann_new_2013}. EOSs of mixtures, therefore, are necessary in addition to EOSs of pure substances.

EOSs of mixtures can be obtained by using the linear mixing approximation (LMA), also known as the additive volume law. LMA states that densities of different materials can be linearly mixed at constant pressure and temperature as
\begin{equation} \label{eq:lma_rho}
    \frac{1}{\rho_{\rm LMA} (P, T)} = \sum_{i=1}^N \frac{x_i}{\rho_i (P,T)},
\end{equation}
where $\rho_{\rm LMA}$ is density of the mixture, $x_i$ denotes the mass fraction of each component, and $\rho_i$ the density of each material. 

The validity of LMA is well tested. LMA is commonly adopted for calculating the EOS of hydrogen-helium mixtures, with modest error of the order of a few percent \citep[e.g.,][]{chabrier_new_2019}. For icy mixtures consisting of H$_2$O, CH$_4$, and NH$_3$, \cite{bethkenhagen_planetary_2017} found that LMA differs from an ab initio simulation by only $\sim4\%$. For H$_2$O-MgSiO$_3$ mixture, LMA overpredicts density by 0.3\% compared to an ab initio simulation at 7,000 K \citep{kovacevic_miscibility_2022}. To probe a wide range of Uranus interior compositions, we calculate the EOS of H$_2$O-CH$_4$-NH$_3$-H/He mixture using LMA based on pure CH$_4$ and pure NH$_3$ EOSs presented in \cite{bethkenhagen_planetary_2017}. The \cite{bethkenhagen_planetary_2017} EOSs for CH$_4$ and NH$_3$ were simulated using density functional theory molecular dynamics (DFT-MD) along a 2,000 K isotherm, representative of Uranus' interior temperature. The mixed-composition ice layer assumes a mixing fraction of 4:1:7 of C:N:O, resembling the solar elemental abundance \citep{asplund_chemical_2009}.

\subsubsection{Temperature Profile} \label{sec:temp_profile}
Here we describe how the temperature profile of each layer is modelled in \texttt{CORGI}. For the Fe and MgSiO$_3$ layers, an isothermal temperature profile is assumed by default because their densities have negligible dependence on temperature. Optionally, the user can calculate an adiabatic temperature profile assuming the core and mantle are a single convecting layer \citep[e.g.,][]{valencia_internal_2006, boujibar_superearth_2020}. Realistically, the iron-silicate part of a planet is not fully adiabatic, but has both conductive and convective regions. Therefore, \texttt{CORGI} also implements the thermal model proposed by \cite{wagner_rocky_2012} based on mixing length theory that applies to both conductive and convective parts.

We assume that the H/He envelope and the gaseous part of the H$_2$O layer, if present, are separated into a lower optically thick convective part and an upper optically thin radiative part by a radiative-convective boundary (RCB), which is defined by the onset of convective instabilities \citep[following][]{rogers_framework_2010}. In the convective part, we assume an adiabatic temperature profile
\begin{equation}
    \left. \left( \frac{\partial \ln T}{\partial \ln P} \right)  \right|_S = \nabla_{\rm ad},
\end{equation}
where $\nabla_{\rm ad}$ is the adiabatic gradient supplied in EOS tables \citep{haldemann_aqua_2020, chabrier_new_2019, chabrier_new_2021}.

Above the RCB, in the radiative part of the atmosphere, an analytical temperature profile is the most appropriate \citep[following][]{guillot_radiative_2010}:
\begin{equation} \label{eq:guillot_t_profile}
\begin{split}
    T^4 = &\frac{3T_{\rm int}^4}{4} \left( \frac{2}{3} + \tau \right) + \\
    &\frac{3T_{\rm irr}^4}{4} f \left[ \frac{2}{3} + \frac{1}{\gamma \sqrt{3}} + \left( \frac{\gamma}{\sqrt{3}} - \frac{1}{\gamma \sqrt{3}} \right) e^{-\gamma \tau \sqrt{3}} \right],
\end{split}
\end{equation}
where $\sigma T_{\rm irr}^4 = f T_{\rm eq}^4$ is the flux received from the host star, $f$ is the redistribution factor, $\sigma T_{\rm int}^4$ is the planet's intrinsic heat flux, and $\gamma$ is the ratio of visible to infrared opacities. We adopt a fiducial value of $\gamma=1$ and assume $f=1/2$ \citep[following][]{rogers_framework_2010}. The optical depth, $\tau$, can be solved from
\begin{equation}
    \frac{d\tau(r)}{dm} = - \frac{\kappa}{4 \pi r^2},
\end{equation}
where the opacity $\kappa$ is a function of $P$ and $T$. We adopt the Rosseland mean opacities for H/He tabulated in \cite{freedman_line_2008}.


\subsubsection{Distinct-layer Model Validation} \label{sec:distinct_layer_validation}
We validate the distinct-layer forward model of \texttt{CORGI} by comparing it to similar models in the literature. 
We generated three sets of planet models representing three categories of planets: on Earth-composition model for Earth-sized terrestrial planets, three compositional K2-18 b models for sub-Neptunes, and two compositional models for Uranus and Neptune.

\begin{figure}
    \centering
    \includegraphics[width=\linewidth]{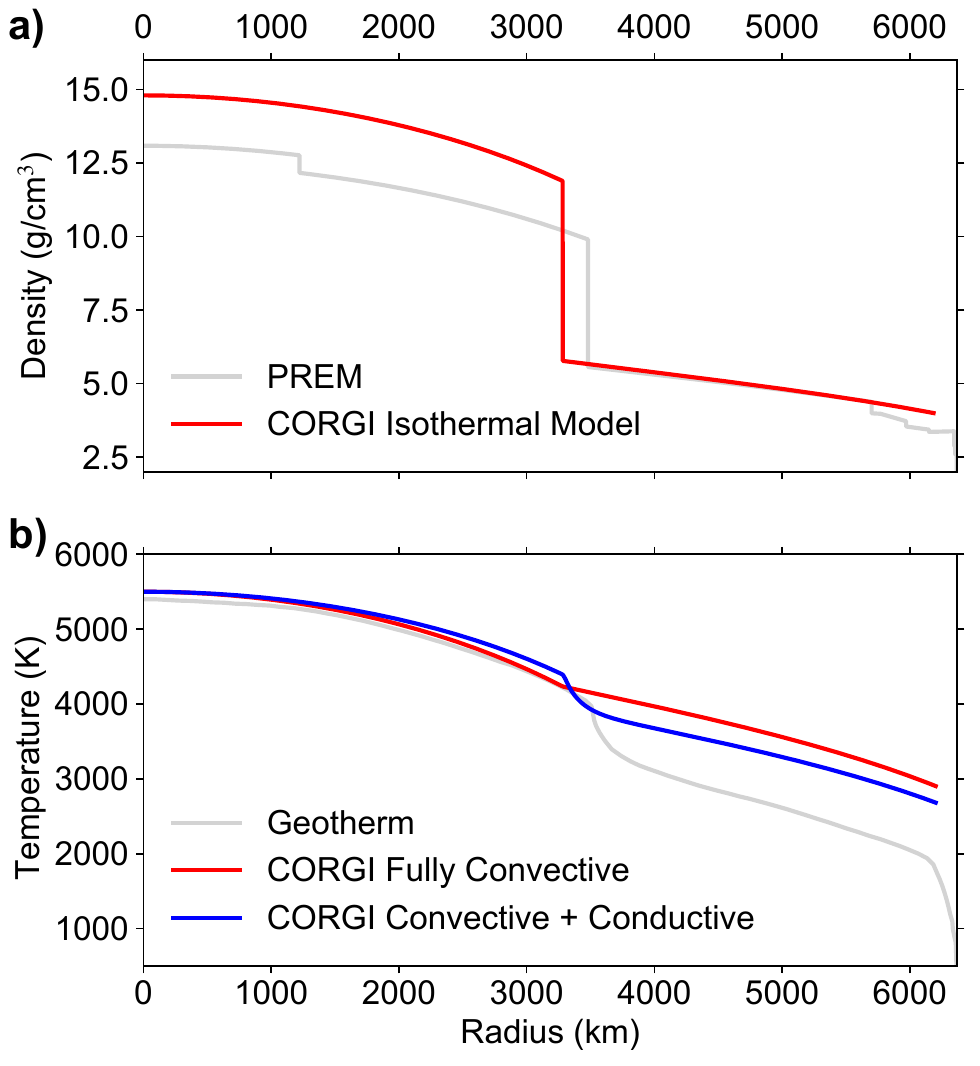}
    \caption{Validation of Earth interior models generated by \texttt{CORGI}. \textbf{a)} $\rho(r)$ profile compared to the PREM. \textbf{b)} $T(r)$ profile compared to the geotherm \citep{hirose_composition_2013}. \texttt{CORGI} model in a) assumes (red) isothermal temperature profile, while models in b) assume (red) a fully convective and adiabatic interior and (blue) a mixing length theory temperature profile with both convective and conductive components \citep{wagner_rocky_2012}. Both the $\rho(r)$ and $T(r)$ profiles display reasonable agreement with measured Earth profiles, validating our forward interior structure model.}
    \label{fig:earth_validation}
\end{figure}

Our Earth model assumes a 32.5\% iron core mass fraction (CMF), a 67.5\% MgSiO$_3$ mantle mass fraction, and an isothermal temperature profile. The calculated mass and radius are $0.9996\,M_\oplus$ and $0.9708\,R_\oplus$, which differ from the ground truth ($M_\oplus=5.97\times10^{24}$ kg, $R_\oplus=$ 6,371 km) by 0.04\% and 2.92\%, respectively (considerably smaller than typical exoplanet mass and radius uncertainties). The density profile of the \texttt{CORGI} Earth model matches the PREM relatively well (Figure \ref{fig:earth_validation}a), with the core $\approx20\%$ denser (13,138 kg m$^{-3}$ compared to 10,987 kg m$^{-3}$) and the mantle $\approx7\%$ denser (4,779 kg m$^{-3}$ compared to 4,449 kg m$^{-3}$). The core-mantle boundary pressure predicted by the \texttt{CORGI} model is slightly higher than reality (155 GPa compared to 135 GPa) due to the denser mantle. Note that we do not expect a perfect match as our model is oversimplified by ignoring details including light elements in core, iron in the mantle, and phase transitions. In addition, the isothermal approximation neglects the solid-liquid core transition and is more appropriate for super-Earths. For the latter, more of the planet's mass is highly compressed, reducing thermal effects \citep[e.g.,][]{seager_massradius_2007, rogers_framework_2010}.

To validate thermal models for iron and silicates, we generated two additional Earth models assuming adiabatic (fully convective core and mantle) and mixing length theory (both convection and conduction regions are present) temperature profiles, and compare them to the geotherm (Figure \ref{fig:earth_validation}b). The adiabatic models are in good agreement with the geotherm in the core. The conductive layer near the core-mantle boundary, however, shows a less steep gradient in our model than in reality. While matching the geotherm in mantle is possible by fine-tuning mixing length theory parameters, here we assume some fiducial parameters following \cite{wagner_rocky_2012} for generality, because such parameters are unknown for the deep interiors of Uranus, Neptune, or exoplanets.

\begin{figure*}
    \centering
    \includegraphics[width=\linewidth]{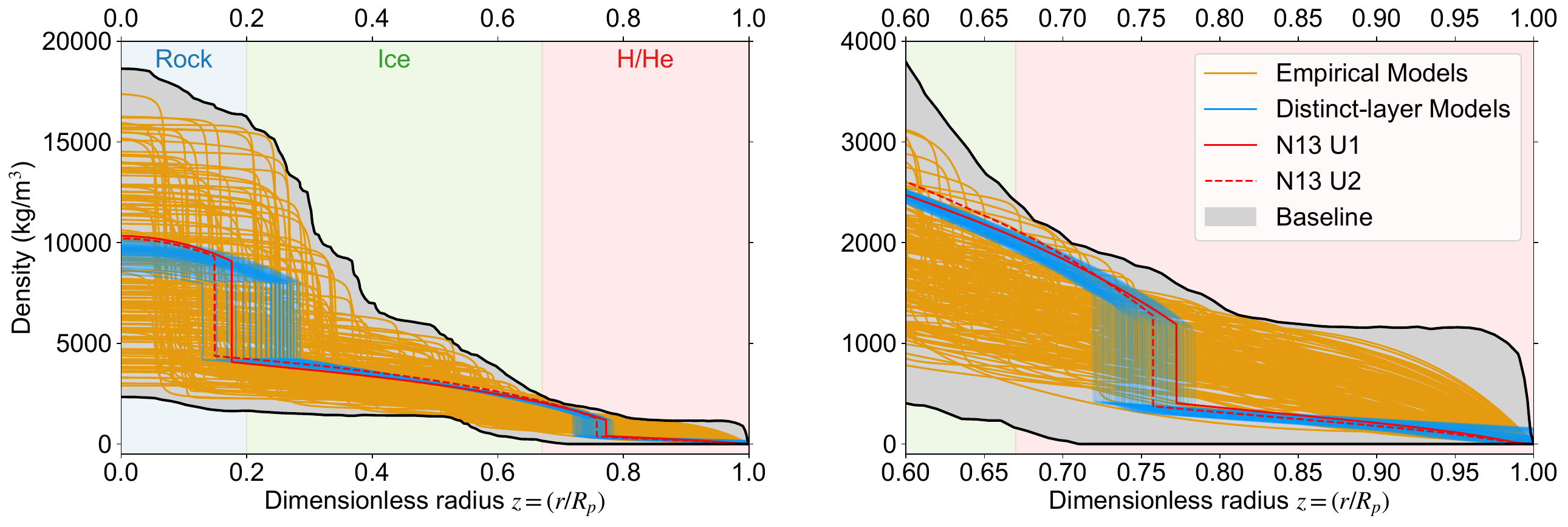}
    \caption{Density profiles from \texttt{CORGI}. Shown are density of all forward models as a function of scaled radius ($z \equiv r/R_p$) in two different ranges: \textbf{a)} the entire planet from $z=0$ to $z=1$ and \textbf{b)} a zoomed-in view of the atmosphere from $z=0.6$ to $z=1$. The gray shaded area with black contours is the baseline region defined in \cite{movshovitz_promise_2022}, which is spanned by all $\rho(z)$ profiles that integrate to the correct total mass. The orange curves are empirical $\rho(z)$ profiles of empirical density models. The cyan curves are $\rho(z)$ profiles of SPI mantle models generated by the distinct-layer method. The red solid and red dashed curves are two distinct-layer Uranus models from \cite{nettelmann_new_2013} for comparison. Background colors represent the approximate radius ranges of rocky core (blue), icy mantle (green), and H/He envelope (red) for the distinct-layer structure. The empirical density models span a much wider parameter space than the distinct-layer models.}
    \label{fig:RGIG_rhoz_baseline}
\end{figure*}

To test the performance of \texttt{CORGI} for sub-Neptunes, we generated three models with drastically different compositions for the extrasolar temperate sub-Neptune K2-18 b following \cite{madhusudhan_interior_2020}.
Model 1 is an extremely iron-rich super-Earth with a thick H/He atmosphere assuming $x_{\rm Fe} = 94.7\%$, $x_{\rm H_2O} = 0.3\%$, and $x_{\rm H/He} = 5\%$.
Model 2 is a sub-Neptune assuming $x_{\rm Fe} = 14.85\%$, $x_{\rm MgSiO_3} = 30.15\%$, $x_{\rm H_2O} = 54.97\%$, and $x_{\rm H/He} = 0.03\%$.
Model 3 is a water world assuming $x_{\rm Fe} = 3.3\%$, $x_{\rm MgSiO_3} = 6.7\%$, $x_{\rm H_2O} = 89.994\%$, and $x_{\rm H/He} = 0.006\%$.
All models assume an internal temperature of 50 K, isothermal core and mantle, an adiabatic water layer, and an H/He atmosphere with both convective and radiative parts. The calculated mass and radius deviations from the observed values ($M_p = 8.63\pm1.35 M_\oplus$, $R_p = 2.610\pm0.087 R_\oplus$, \citealt{Benneke_water_2019}) are $\Delta M/M = 0.16\%$ and $\Delta R/R = 5.51\%$ for model 1, $\Delta M/M = 0.01\%$ and $\Delta R/R = 1.45\%$ for model 2, and $\Delta M/M < 10^{-7}$ and $\Delta R/R = 2.72\%$ for model 3. These are generally smaller than the observational uncertainties, validating our models. Model 1, which has the thickest H/He envelope (5\% by mass), displays the largest deviation, which is due to the sensitive dependence of the H/He envelope on assumptions such as internal temperature and mean molecular weight.

Our Uranus and Neptune models are compared to two Uranus models (hereafter N13 U1 and N13 U2) and three Neptune models presented in \cite{nettelmann_new_2013}, all of which assume a distinct-layer structure. General agreement between our models and theirs are observed, further validating \texttt{CORGI}. See Section \ref{sec:results} for detailed comparisons.

In summary, \texttt{CORGI} is capable of modeling the interior structures of planets with diverse compositions and sizes, with sufficient precision to apply to Uranus, Neptune, and exoplanets.


\subsection{Empirical Density Forward Model} \label{sec:empirical_density_model} 
Unlike distinct-layer models constructed from first principles, empirical density models generate monotonic functions of density, $\rho(r)$, that match a planet's mass and radius without considering the physical reality of such density profiles. While it is possible to generate mixture EOSs using LMA to produce smooth density profiles without layer boundaries, it is generally more efficient to use some parametrization to generate $\rho(r)$ functions that are agnostic to the underlying physical composition. Here, we use the parametrization introduced in \cite{movshovitz_promise_2022}, which represents $\rho(z)$, where $z \equiv r/R_p$ is the dimensionless radius, as a continuous and continuously differentiable function
\begin{equation} \label{eq:movshovitz_rhoz_equation}
\begin{split}
    \rho(z) = &\rho_0 + \sum_{n=2}^8 a_n (z_n - 1) + \\
    &\sum_{n=1}^2 \frac{\sigma_n}{\pi} \left[\frac{\pi}{2} + \arctan(-\nu_n (z-z_n)) \right],
\end{split}
\end{equation}
where $\rho(z=1) = \rho_0$ is the surface density, generally represented by density at 1 bar. The first sum represents an eighth degree polynomial with user-defined coefficients, $a_n$. The second sum defines two density jumps at locations $z_1$ and $z_2$, with the sharpness of the jumps controlled by $\nu_1$ and $\nu_2$ and the height of the jumps controlled by $\sigma_1$ and $\sigma_2$. We use the above parametrization to generate $\sim100$ empirical density models that integrate to the correct total mass of Uranus (Figure \ref{fig:RGIG_rhoz_baseline}).

Unlike for distinct layer models, we do not run interior retrievals to estimate empirical density models. This is because Equation (\ref{eq:movshovitz_rhoz_equation}) is computationally inexpensive. We use a brute force approach to generate millions of profiles with random $a_n$, $z_n$, $\sigma_n$, and $\nu_n$, to find $\rho(z)$ profiles that integrate to the correct mass, instead of using the MCMC sampler.

\subsection{Planet Interior Retrieval} \label{sec:inverse_model}
While the distinct-layer forward modeling module of \texttt{CORGI} can efficiently solve for the mass and radius of a hypothetical planet given $P_{\rm c}$ and mass fractions of each layer $\{x_i\}$, the inverse problem -- solving for internal structure given mass and radius -- is the more critical one. To solve the inverse problem assuming a distinct layer structure, we developed an interior retrieval program based on \texttt{emcee}, a \texttt{Python} implementation of the Markov chain Monte Carlo (MCMC) ensemble sampler \citep{foreman-mackey_emcee_2013, foreman-mackey_emcee_2019}. A preliminary version of this retrieval code has been applied to solve for the possible interior compositions of recently discovered super-Earth TOI-1075 b \citep{essack_toi-1075_2023}.

The interior retrieval priors are defined as follows. Uranus has a mass of $14.54\,M_\oplus$, a volumetric mean radius of $3.981\,R_\oplus$,
and a mean temperature of $T_{\rm mean} = 76$ K at 1 bar.\footnote{\href{https://nssdc.gsfc.nasa.gov/planetary/factsheet/}{https://nssdc.gsfc.nasa.gov/planetary/factsheet/}} We assume Uranus has a three-layer structure with a MgSiO$_3$ core, an icy mantle, and an H/He-dominated envelope. 

We explore the following compositions for the mantle and the envelope:
\begin{enumerate}
    \item Mixed H$_2$O-H/He. For this composition, both the mantle and the atmosphere are assumed to be composed of mixture of H$_2$O and H/He. The ratios at which H$_2$O and H/He are mixed are defined by the heavy element mass fraction, $Z$. The heavy element mass fractions in both the mantle ($Z_{\rm mantle}$) and the atmosphere ($Z_{\rm atm}$) are allowed to vary between 0 and 1. This composition is similar to that assumed in \cite{nettelmann_new_2013}, which found generally high $Z_{\rm mantle}$ values (0.915 and 0.944 for U1 and U2 models in \citealt{nettelmann_new_2013} models, respectively), and generally low $Z_{\rm atm}$ values (0.17 and 0.08 from the same models).
    \item Synthetic planetary ice \citep[SPI; see e.g.,][]{bethkenhagen_planetary_2017, guarguaglini_laser-driven_2019} mantle and H/He-dominated atmosphere. The mantle is assumed to be composed of H$_2$O-CH$_4$-NH$_3$ mixture ice, where the C:N:O elemental fraction is assumed to be 4:1:7, resembling the solar abundance \citep{asplund_chemical_2009}. Light elements (H/He) are allowed to mix into the SPI mantle, and heavy elements (H$_2$O-CH$_4$-NH$_3$) are allowed to mix into the atmosphere. The following $Z$ values are assumed: $Z_{\rm atm}$ = 12.5\% or 25\%, and $Z_{\rm mantle}$ = 85\%, 95\%, or 100\%, totaling 6 pairs.
    \item Pure H$_2$O mantle and pure H/He envelope. Although this composition is an oversimplification for Uranus, it is a common assumption when modeling the interior structure of exoplanets and is a useful reference case.
\end{enumerate}
In all mixture models above, density is calculated using the LMA, while temperature is assumed to follow the pure H$_2$O or pure H/He adiabat.

An iron core is not included in our distinct-layer Uranus models. This is because for planets formed beyond the water ice line, Fe is expected to be fully oxidized and mixed with other rocky materials, rather than being segregated into a metallic core \citep{vazan_new_2022}. In any case, iron in the rocky layer leads to a negligible density increase relative to pure MgSiO$_3$.

There are six (the last two only apply to the mixed H$_2$O-H/He models) free parameters: CMF, $x_{\rm Z}$, $P_{\rm c}$, $T_{\rm c}$, $Z_{\rm mantle}$, and $Z_{\rm atm}$. CMF defines the mass of the rocky core relative to the combined mass of core plus icy mantle. $x_{\rm Z}$ defines the mass fraction of core plus icy mantle relative to the total planetary mass. Therefore, $x_{\rm core} = {\rm CMF} \cdot x_{\rm Z}$, $x_{\rm mantle} = (1-{\rm CMF}) \cdot x_{\rm Z}$, and $x_{\rm atm} = 1 - x_{\rm core} - x_{\rm mantle}$. We define the mass fractions of layers indirectly via CMF and $x_{\rm Z}$ to reduce coupling between parameters, improving the efficiency of MCMC sampler.
Given the expected small core mass and large ice mass fraction of the distinct-layer structure \citep[e.g.,][]{nettelmann_new_2013}, the prior for CMF is a uniform distribution between 0 and 0.25 and the prior for $x_{\rm Z}$ is a uniform distribution between 0.7 and 1. The prior for $P_{\rm c}$ is a log-normal distribution centered at $10^{12}$ Pa, or 1000 GPa. The prior for $T_{\rm c}$ is a uniform distribution between 300 K and 15,000 K. The prior for $Z_{\rm mantle}$ is a uniform distribution between 0.85 and 1. The prior for $Z_{\rm atm}$ is a uniform distribution between 0.05 and 0.2. Note that these assumed priors do not introduce bias, because a MCMC retrieval run with large enough step numbers is agnostic of initial conditions. The retrieval model was run with 2,000 walkers for more than 1,000 steps, with convergence reached around 800 steps.

\begin{figure}
    \centering
    \includegraphics[width=\linewidth]{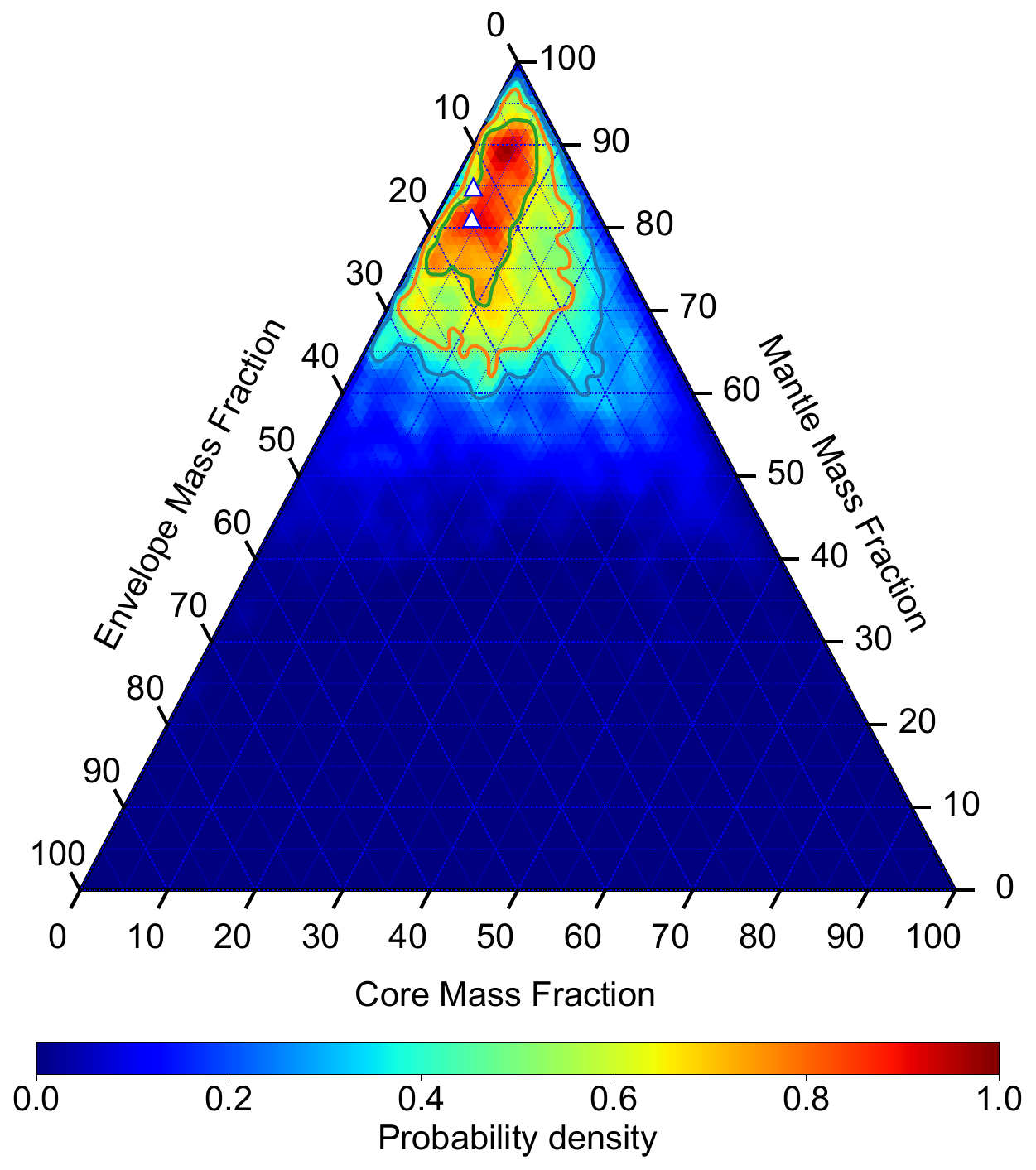}
    \caption{Ternary diagram showing the retrieved probability distribution of the interior composition of Uranus, assuming mixed H$_2$O-H/He composition. The best-fit interior composition is $x_{\rm core} = 0.11^{+0.14}_{-0.07}$, $x_{\rm mantle} = 0.71^{+0.13}_{-0.14}$, $x_{\rm atm} = 0.14^{+0.12}_{-0.09}$. The most probable compositions are outlined by contours enclosing (from innermost to outermost, colored green, red, and blue, respectively) 30\%, 50\%, and 68\% of all data points in the densest region. The two Uranus models presented in \cite{nettelmann_new_2013} are shown as triangle markers (upper marker for U2 and lower marker for U1) and fall within the retrieved 30\% contours.}
    \label{fig:uranus_retrieval}
\end{figure}

The interior retrieval results for Uranus are shown in a ternary diagram (Figure \ref{fig:uranus_retrieval}). The mixed H$_2$O-H/He model is shown because it assumes the same interior composition and structure as previous works \citep[e.g.,][]{fortney_interior_2010, nettelmann_new_2013}, providing the best ground for comparison. Results for other compositional models are summarized in Appendix \ref{sec:appendix_corner_plot}. The best-fit interior composition is $x_{\rm core} = 0.11^{+0.14}_{-0.07}$, $x_{\rm mantle} = 0.71^{+0.13}_{-0.14}$, $x_{\rm atm} = 0.14^{+0.12}_{-0.09}$, $Z_{\rm mantle} = 0.85^{+0.10}_{-0.15}$, and $Z_{\rm atm} = 0.28^{+0.30}_{-0.18}$. These layer mass fractions are consistent with previous study by \cite{nettelmann_new_2013} assuming a similar three-layer structure. Their U1 model has $x_{\rm U1, core} = 0.042$, $x_{\rm U1, mantle} = 0.811$, $x_{\rm U1, H/He} = 0.147$, $Z_{\rm U1, mantle} = 0.915$, and $Z_{\rm U1, atm} = 0.17$. Their U2 model has $x_{\rm U2, core} = 0.025$, $x_{\rm U2, mantle} = 0.849$, $x_{\rm U2, H/He} = 0.126$, $Z_{\rm U2, mantle} = 0.944$, and $Z_{\rm U2, atm} = 0.08$. Both U1 and U2 models reside within the innermost contour, which encloses 30\% of all data points, of the posterior compositional distribution (Figure \ref{fig:uranus_retrieval}). Their $Z$ values are within error with our retrieval results, except for $Z_{\rm U2, atm}$. The best-fit central pressure is $\log_{10}(P_{\rm c}) = 11.97^{+0.18}_{-0.19}$, or approximately $933^{+479}_{-331}$ GPa. The estimated $T_{\rm c}$ has a wide dispersion of $3411^{+2032}_{-1939}$ K. We note that this $T_{\rm c}$ range is lower than realistic values and is a retrieval artifact. Distinct-layer models assuming an adiabatic temperature profile generally have core temperatures $\sim6000$ K \citep[e.g.,][]{nettelmann_new_2013}. The peak at low temperatures (see Appendix \ref{sec:appendix_corner_plot}) represents isothermal models that happen to fit the mass and radius constraints but are not physical. Therefore, our distinct-layer forward models assume a higher $T_{\rm c}$ of 5500 K than the retrieved value.

We perform the same retrieval analysis for all distinct-layer mantle and atmosphere compositions. Within the 50\% best-fit compositional contour of each compositional scenario (e.g., red contour, Figure \ref{fig:uranus_retrieval}), we select $\sim100$ evenly distributed combinations of layer mass fractions to generate $\rho(r)$ profiles using our forward interior structure module (blue lines, Figure \ref{fig:RGIG_rhoz_baseline}) and simulate their gravity harmonics using the CMS module, which will be introduced in detail in the next subsection. The choice of best-fit contour that encloses 50\% of all data points is somewhat arbitrary, but is justified because this contour outlines a wider compositional parameter space than the retrieved $1\,\sigma$ ranges (i.e., the $x_i$ values we report here, see also corner plot in Appendix \ref{sec:appendix_corner_plot}). Exploring a wide compositional parameter space is beneficial because we are interested in the maximally possible range of $J_n$ that can be produced by a certain mantle and atmosphere composition.

\subsection{Concentric Maclaurin Spheroid (CMS) Model} \label{sec:CMS_model_details}
Gravity measured through precise Doppler tracking of spacecraft in circumplanetary orbits is the best method for accurately determining the interior density distribution of a planet from space \citep[e.g.,][]{iess_measurement_2018, iess_measurement_2019, militzer_models_2019, militzer_juno_2022}. To make predictions for planets without spacecraft measurements and to interpret the data for planets with those measurements, theoretical models are required. Conventionally, the gravity harmonics of a planet in hydrostatic equilibrium undergoing uniform rotation with rate $\omega$ are obtained by the theory of figures \citep{Zharkov_1978_physics}. The theory of figures was applied to the interpretation of Juno \citep{nettelmann_low-_2017} and Cassini \citep{ni_understanding_2020} measurements, and was recently calculated to the seventh \citep{nettelmann_theory_2021} and the tenth \citep{morf_interior_2024} order.

Alternatively, the CMS method can simulate self-consistent shape and gravity harmonics of a planet to higher $J_n$ terms with improved precision \citep{hubbard_concentric_2013}. The CMS method models a planet as many constant-density spheroids and numerically solves for the gravity field with high numerical precision by using Gaussian quadrature. The model is set up in a way that density discontinuities associated with the distinct-layer structure can be trivially incorporated \citep{hubbard_concentric_2013}. Smooth density transitions of the empirical models can also be approximated with a large enough number of layers. The CMS method has been validated by comparing to an independent consistent level curve (CLC) method and a Bessel function method \citep{wisdom_differential_2016}. Here, we implement the CMS model following the formalism described in \cite{hubbard_concentric_2013} and \cite{militzer_models_2019}.

Note that when ignoring interior dynamics such as deep atmospheric winds, the gravity field of a planet is axisymmetric and north-south symmetric. Therefore, we only model longitude-independent zonal gravity harmonics with even $n$, which dominate the gravity field.

We present a qualitative overview of the CMS method here and present the detailed formalism in Appendix \ref{sec:CMS_details}. The planet is divided into $N_L$ layers indexed as $i=0, 1, ..., N_L-1$ with each layer having $N_M$ Gaussian quadrature points indexed as $m=1, 2, ..., N_M$. Because north-south symmetry is assumed, only points in one hemisphere needs to be computed. Initially, the planet is a perfect sphere with constant radius on each layer surface. With this shape parametrization, we guess some arbitrary initial $J_n$, and calculate the gravitational potential, $V$, and centrifugal potential, $Q$, on each grid point given the uniform rotation rate $\omega$. The total potential $U = V + Q$ on each layer surface is now obviously not constant, because constant-potential surfaces in a rotating body in hydrostatic equilibrium should be oblate spheroids. One may use a Newton step to minimize the potential difference on each grid point with regard to a reference point, chosen to be a point on the layer's equator. With this updated shape, we calculate new $J_n$ values and then use these $J_n$'s to update the shape of the planet (see details in Appendix \ref{sec:CMS_details}). This process is iterated until the difference in $J_n$ estimates between consecutive iterations, $\Delta J_n$, falls below some small tolerance ($\lesssim 10^{-12}$, see Appendix \ref{sec:appendix_CMS_validation}).

We validate our CMS model by comparison to two independent methods, namely the CLC method and a method applying spherical Bessel functions \citep{wisdom_differential_2016}. We present detailed model validation and discuss the convergence of our CMS model in Appendix \ref{sec:appendix_CMS_validation}.


\section{Results} \label{sec:results}
\begin{figure*}
    \centering
    \includegraphics[width=\linewidth]{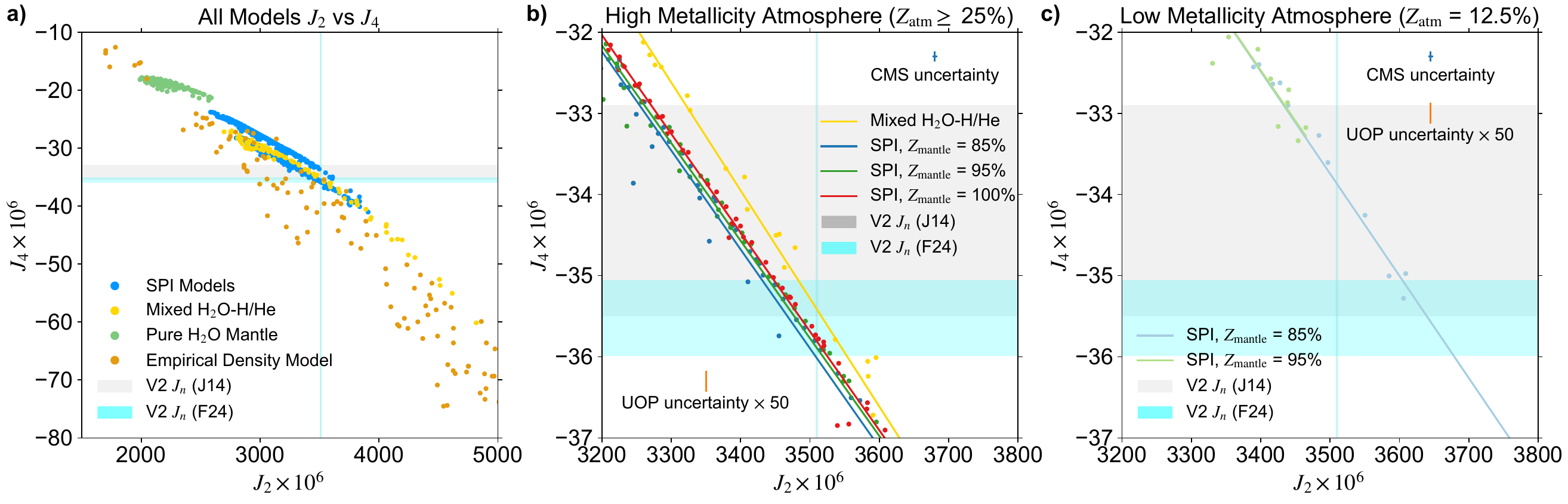}
    \caption{CMS model results for $J_2$ and $J_4$. \textbf{a)} All compositional and structural models overplotted with V2 measurements. Two V2 data analyses are plotted: (gray) J14 (\citealt{jacobson_orbits_2014}) and (cyan) a recent reanalysis by F24 (\citealt{french_uranus_2024}). \textbf{b)} Distinct-layer models with high atmospheric metallicities ($Z_{\rm atm} \geq 25\%$). Lines represent linear fit to CMS $J_n$ results. CMS simulation uncertainty (blue) and predicted UOP measurement uncertainty $\times50$ from \cite{parisi_uranus_2024} (orange) are plotted for comparison. \textbf{c)} Distinct-layer models with low atmospheric metallicities ($Z_{\rm atm} = 12.5\%$). Pure H$_2$O mantle models are confidently ruled out, while mixed H$_2$O-H/He, SPI, and empirical density models are consistent with V2 measurements. The high $Z_{\rm atm}$ models are consistent with the F24 analysis with smaller $J_4$, while the low $Z_{\rm atm}$ models are consistent with the J14 analysis. Future UOP measurements of $J_2$ and $J_4$, with much smaller uncertainties than V2, will rule out one category of atmospheric metallicity model. SPI models with higher levels of mixing are more consistent with V2 data (see Section \ref{sec:high_mixing_j2j4}).
    }
    \label{fig:j2j4_error_bar}
\end{figure*}

Here, we summarize results derived from \texttt{CORGI} interior and gravity field modeling.

\subsection{High Levels of Mixing Are Required to Explain Voyager 2 $J_2$ and $J_4$ Measurements} \label{sec:high_mixing_j2j4}

Our major finding is that only cases with high levels of mixing can explain the V2 measurements of Uranus' $J_2$ and $J_4$. Highly mixed models include both empirical density models, which by definition can be arbitrarily mixed, and distinct-layer models with relatively high H/He mass fraction in the icy mantle and relatively high heavy element mass fraction in the envelope. Quantitatively, to be consistent with V2 measured zonal gravity harmonics of Uranus, distinct-layer models either need to have high atmospheric metallicity ($Z_{\rm atm} \geq 25\%$, Figure \ref{fig:j2j4_error_bar}b), or have low $Z_{\rm mantle}$ on the order of 85\% (Figure \ref{fig:j2j4_error_bar}c). Our results confirm that empirical models \citep[e.g.,][]{movshovitz_promise_2022} and distinct-layer models with mixed H$_2$O-H/He interiors \citep[e.g.,][]{nettelmann_new_2013} can reproduce V2 $J_2$ and $J_4$, while providing the novel insight that some SPI models with high levels of mixing can reproduce the measurements as well.

\texttt{CORGI} results offer the first detailed gravity harmonics constraints for Uranus interiors with synthetic planetary ice. We show that SPI models with low levels of mixing -- low atmospheric metallicity ($Z_{\rm atm}=12.5\%$) and high mantle heavy element mass fraction ($Z_{\rm mantle}=95\%$ or 100\%) -- can be robustly ruled out by $J_2$ and $J_4$ measurements (Figure \ref{fig:j2j4_error_bar}c, in which $Z_{\rm mantle}=100\%$ models are outside of the plotting region).

\begin{figure*}
    \centering
    \includegraphics[width=\linewidth]{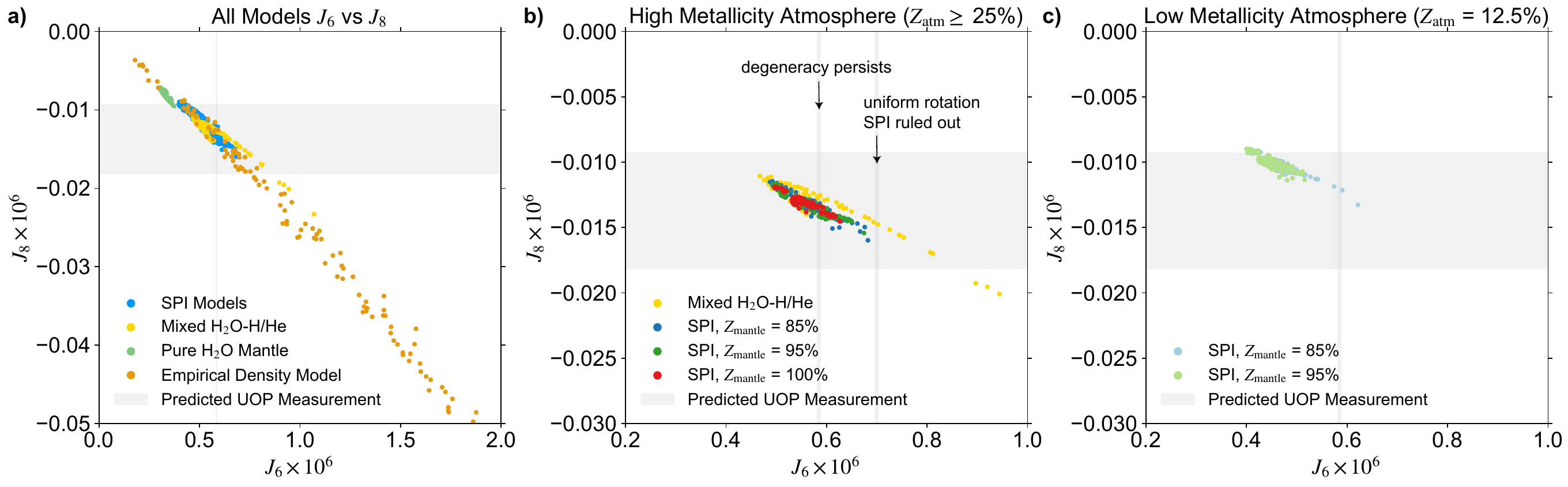}
    \caption{CMS model results for $J_6$ and $J_8$. \textbf{a)} All compositional and structural models overplotted with predicted UOP measurements. Width of the gray line/region represents predicted UOP uncertainty assuming close-in inclined polar orbit \citep{parisi_uranus_2024}. \textbf{b)} Distinct-layer models with high atmospheric metallicities ($Z_{\rm atm} \geq 25\%$). Two hypothetical $J_6$ measurements are shown (indicated by arrows with annotations). \textbf{c)} Distinct-layer models with low atmospheric metallicities ($Z_{\rm atm} = 12.5\%$). UOP $J_6$ measurement can potentially rule out some uniformly rotating distinct-layer models, thanks to its small uncertainty. UOP $J_8$ measurement will offer limited information due to large uncertainty. 
    }
    \label{fig:j6j8_error_bar}
\end{figure*}

Mixed H$_2$O-H/He cases show excellent agreement with V2 $J_2$ and $J_4$ measurements (yellow points and line, Figure \ref{fig:j2j4_error_bar}b), compatible with both data analyses. This result agrees with previous literature \citep{nettelmann_new_2013}.

Empirical density models, which allow arbitrary compositional gradients and therefore naturally have high levels of mixing, span an extensive parameter space that completely eclipse distinct-layer models and are fully consistent with Uranus' measured zonal gravity harmonics (orange points, Figure \ref{fig:j2j4_error_bar}a). Because empirical density models are generated with parametrization \citep{movshovitz_promise_2022}, which contain no information on the mass fractions of planetary building blocks, we can only conclude that significant mixing is present but cannot provide quantitative constraints on mass fractions of planet-building materials inside the mixed interior. Future studies that properly handle the EOS, transport properties, and P-T profile of a mixed interior are required to provide precise mass fractions in the interiors of empirical density models.

\subsection{UOP $J_2$ and $J_4$ Measurements Will Reveal Atmospheric Metallicity}

Another implcation of \texttt{CORGI} results is that with V2 data, high and low atmospheric metallicity ($Z_{\rm atm}$) models cannot be distinguished, while future UOP $J_2$ and $J_4$ measurements will allow such distinction. Comparison between our CMS results and V2 measurements reveal that all models, except for the pure H$_2$O mantle model that is confidently ruled out, are consistent with V2 $J_2$ and $J_4$ (Figure \ref{fig:j2j4_error_bar}a). Empirical models cover a wide parameter space, eclipsing all distinct-layer models, as expected from $\rho(z)$ profiles (Figure \ref{fig:RGIG_rhoz_baseline}). Distinct-layer models can be separated into two categories based on $Z_{\rm atm}$. The high metallicity category ($Z_{\rm atm} = 25\%$ for the SPI models and $Z_{\rm atm}=28\%$ for the mixed H$_2$O-H/He models) has lower $J_4$ and are consistent with the recent \cite{french_uranus_2024} reanalysis of V2 data (Figure \ref{fig:j2j4_error_bar}b). The low metallicity category ($Z_{\rm atm} = 12.5\%$ SPI models) has higher $J_4$ and agrees with the \cite{jacobson_orbits_2014} analysis of V2 data (Figure \ref{fig:j2j4_error_bar}c).

Due to the large uncertainties in V2 $J_4$ measurement and data analysis, we cannot conclude which distinct-layer model is a better fit for the interior of Uranus. With the much reduced uncertainty UOP will offer (orange error bar, which is magnified by 100 times, Figure \ref{fig:j2j4_error_bar}b and c), such compositional degeneracy can be reduced. Predicted UOP $J_4$ uncertainty in an inclined polar orbit with periapsis within the rings \citep{parisi_uranus_2024} is much smaller than the $J_4$ difference between high and low atmospheric metallicity models. With UOP gravity harmonics measurements, one $Z_{\rm atm}$ category will be confidently ruled out. Further constraints on the exact $Z$ values will be limited by the \textit{intrinsic} compositional degeneracy, represented by the scattering of points in Figure \ref{fig:j2j4_error_bar}b and c. Due to this intrinsic degeneracy, differentiating between high $Z_{\rm atm}$ models will have to rely on higher order $J_n$.

\subsection{$J_6$ Measurement by UOP May Rule Out Some Synthetic Planetary Ice Models}
Event though UOP $J_2$ and $J_4$ measurements can potentially distinguish between high and low $Z_{\rm atm}$ scenarios, they are not sufficient. Because high $Z_{\rm atm}$ models are closely clustered together (Figure \ref{fig:j2j4_error_bar}b) and due to the intrinsic compositional degeneracy, we need higher order gravity harmonics to further improve the constraints on the interior of Uranus.

\begin{figure*}
    \centering
    \includegraphics[width=0.95\linewidth]{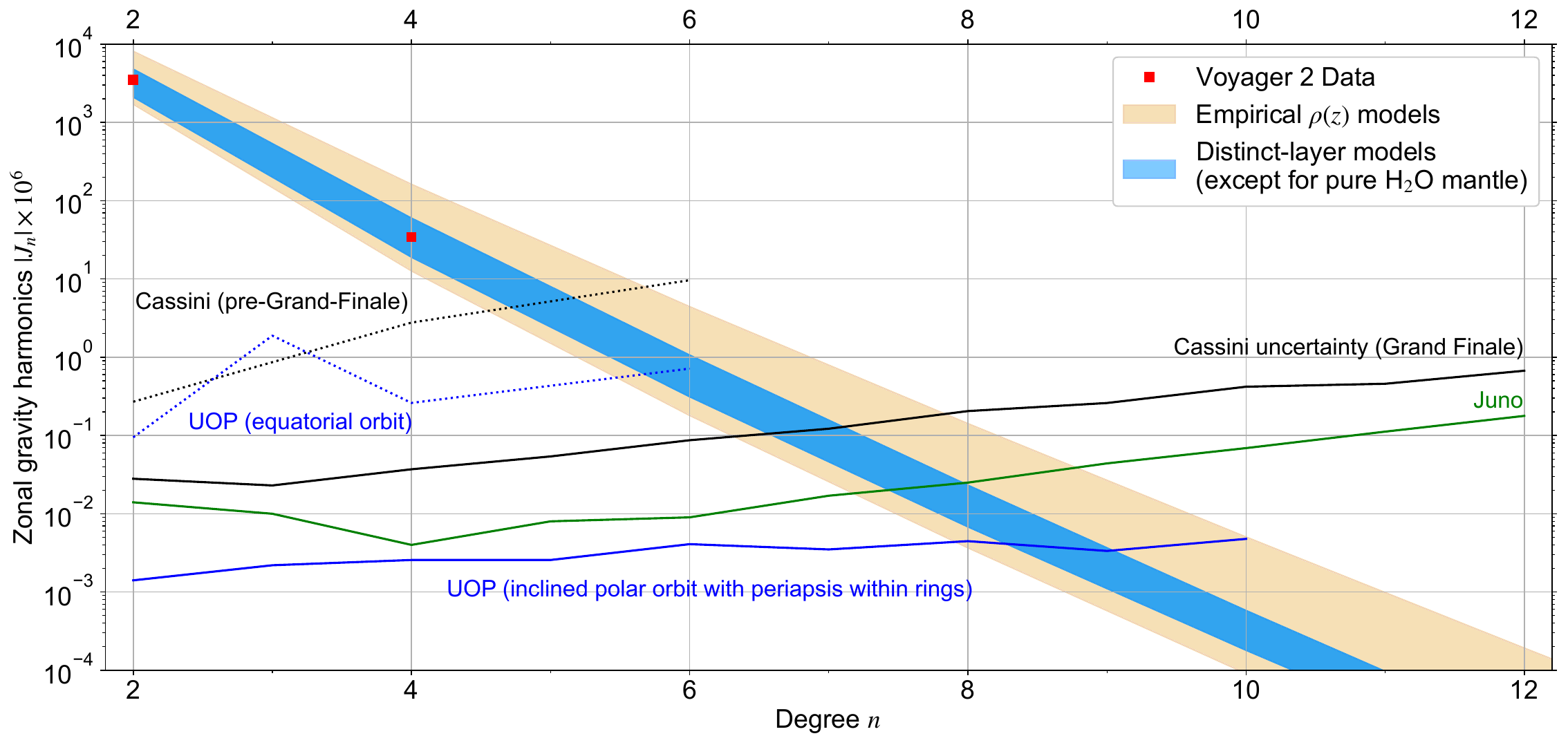}
    \caption{Simulated $J_n$ gravity harmonics as a function of degree $n$. Spacecraft measurement uncertainties are plotted for comparison. The light orange and blue shaded regions show $J_n$ parameter spaces spanned by empirical $\rho(z)$ models and distinct-layer models (except for the pure H$_2$O mantle models), respectively. 
    The red squares show $J_2$ and $J_4$ measured by V2. The measurement uncertainties are smaller than the marker size. 
    Juno uncertainties are shown as green solid line \citep{iess_measurement_2018}. Cassini uncertainties during \citep{iess_measurement_2019} and before \citep{jacobson_gravity_2006} the Grand Finale are shown as black solid and dotted lines, respectively. Predicted UOP uncertainties are colored blue, where the blue dotted line assumes equatorial orbits \citep{Mazarico_Gravity_2023} while the blue solid line assumes inclined polar orbits with periapsis inside the rings \citep{parisi_uranus_2024}.
    Close-in polar orbits are more sensitive to zonal gravity harmonics, allowing UOP to robustly measure $J_6$ and possibly $J_8$.}
    \label{fig:jn_region}
\end{figure*}

Measuring $J_6$ with UOP may reduce the compositional degeneracy by ruling out some SPI models, if small uncertainties can be achieved. The uncertainty of a hypothetical UOP measurement of $J_6$ is much smaller than the $J_6$ value itself (Figure \ref{fig:j6j8_error_bar}b), if UOP is in an inclined polar orbit with periapsis inside the rings (\citealt{parisi_uranus_2024}; see also Figure \ref{fig:jn_region}). In an ideal scenario, if a $J_6$ value that significantly deviates from log-linear extrapolation based on V2 $J_2$ and $J_4$ data is observed by the UOP (Figure \ref{fig:j6j8_error_bar}b and c, right vertical line), uniformly rotating SPI models can be ruled out because no such model can be consistent with $J_2$, $J_4$, and $J_6$ measurements at the same time. In such a scenario, $J_6$ provides an additional constraint on the interior composition of Uranus. Less ideally, if a $J_6\times10^6$ value around $0.6$ is observed (Figure \ref{fig:j6j8_error_bar}b and c, left vertical line), all models that are consistent with V2 $J_2$ and $J_4$ measurements will still be consistent with the new $J_6$ measurement, offering no new insights into Uranus' interior composition and structure. Nevertheless, even though we cannot rule out some model categories in this scenario, MCMC studies have shown that a combination of multiple higher order $J_n$ harmonics, even with moderate uncertainties, can place tighter constraints on the $\rho(r)$ profile than using $J_2$ and $J_4$ alone, even with improved uncertainties \citep{movshovitz_promise_2022}.

A $J_8$ measurement by UOP will not help resolve the interior degeneracy unless a significantly smaller uncertainty than predicted can be achieved. Even in an close-in inclined polar orbit, which is ideal for capturing the zonal variations in gravity field, measurement error predicted for $J_8$ is almost as large as the $J_8$ value itself (Figure \ref{fig:j6j8_error_bar} and \ref{fig:jn_region}).

\subsection{The Pure H$_2$O Mantle Assumption is Invalid}
Distinct-layer models with a pure H$_2$O mantle are not consistent with V2 measurements of zonal gravity harmonics of Uranus (Figure \ref{fig:j2j4_error_bar}a). Our simulated $J_2$ ($\times10^6$) values for the pure H$_2$O mantle models range from 1987.9 to 2592.3, which differs from the V2 data ($J_2 \times 10^6 = 3510.7\pm0.7$ or $3509.291\pm0.412$, according to \citealt{jacobson_orbits_2014} and \citealt{french_uranus_2024}, respectively) by $\sim1000\times$ the measurement uncertainty. Our simulated $J_4$ ($\times10^6$) values range from $-17.8$ to $-21.8$, which deviates by $> 8\times$ the measurement uncertainty from the V2 measurement ($J_4 \times 10^6 = -34.2\pm1.3$ or $-35.522\pm0.466$). Such a large inconsistency occurs despite our having swept a large parameter space of possible layer masses using MCMC samplers, implying that a distinct-layer structure with a pure water mantle is indeed inconsistent with the gravity harmonics of Uranus. 

Our results have implications for exoplanet interior modeling. Even though for Uranus and Neptune, more detailed models akin to our mixed H$_2$O-H/He models involving heavy element mixed into the envelope and light elements mixed into the mantle are commonly adopted \citep[e.g.,][]{fortney_interior_2010, nettelmann_new_2013}, for exoplanets, a pure H$_2$O ice layer is a common simplification \citep[e.g.,][]{seager_massradius_2007, rogers_framework_2010, madhusudhan_interior_2020}. Our understanding of planet formation and interior composition are increasingly dominated by exoplanetary studies, both because exoplanets are numerous and because some exoplanets populate mass-radius parameter spaces lacking solar system counterparts (e.g., super-Earths and sub-Neptunes). Here, our results suggest that the commonly adopted pure H$_2$O layer assumption for exoplanets is oversimplified and inconsistent with the gravity field of a realistic planet. Our planetary interior model, \texttt{CORGI}, is capable of modeling mixed-composition layers with arbitrary $Z$ values. Future studies on interior compositions of exoplanets should therefore take mixed-composition ice and atmosphere layers into consideration.

The main reason that a pure H$_2$O mantle is inconsistent with V2 $J_n$ measurements is that pure H$_2$O is too dense. Around 100 GPa, pure H$_2$O is $\sim40\%$ denser than pure CH$_4$ and $\sim20\%$ denser than binary mixtures of water, methane, and ammonia \citep{bethkenhagen_planetary_2017}. If some mechanism can decrease H$_2$O density, making it comparable to the density of SPI and H$_2$O-H/He mixtures, gravity harmonics of Uranus models with pure H$_2$O layers may become more consistent with measurements. One possible mechanism that reduces water layer density is the introduction of thermal boundary layers \citep[e.g.,][]{nettelmann_uranus_2016}. The strikingly low luminosity of Uranus compared to Neptune favors the existence of a thermal boundary \citep[e.g.,][]{vazan_explaining_2020}. However, we find that thermal boundaries are insufficient to explain the density difference between pure H$_2$O models and mixed-composition models alone. We attempted some models with $ T_{\rm c} =$ 8,500 K (3,000 K hotter than distinct-layer models shown in Figure \ref{fig:RGIG_rhoz_baseline}) and a thermal boundary in the upper atmosphere to allow the upper atmospheric temperature to converge with $T_{\rm eq}$ at 1 bar. These hotter models have lower water density, but are still significantly denser than the N13 U1 and U2 models and are inconsistent with the gravity harmonics measured by V2. If a thermal boundary layer indeed exists, it needs to either produce a $>$ 3,000 K temperature increase, or operates jointly with other mechanisms that decrease the water layer density, such as the inclusion of H/He and/or volatile ices.

Our results imply that realistic EOSs of ice and H/He mixtures are important for probing the interior structure of Uranus, Neptune, and extrasolar intermediate-sized planets. The lack of realistic mixture EOSs is one of the main reasons that the oversimplifying assumption of a pure H$_2$O ice layer is widely adopted for modeling exoplanets \citep[e.g.,][]{seager_massradius_2007, rogers_framework_2010, madhusudhan_interior_2020}. Decades from now, when UOP measures the $J_n$ parameters of Uranus, detailed forward models will be required to interpret such measurement results. The accuracy of those models will rely heavily on experimental and ab initio simulation constraints on EOS of planet-forming materials, especially SPI involving H, C, N, and O \citep[e.g.,][]{guarguaglini_electrical_2021}, under the P-T conditions relevant for Uranus (Figure \ref{fig:h2o_pt_aqua}). Currently, experimental constraints on such mixtures are insufficient, and we have to rely on LMA-based mixture EOS \citep[e.g.,][]{bethkenhagen_superionic_2015, bethkenhagen_planetary_2017}. Therefore, we call for more experiments and simulations on the physical properties of realistic planet-building mixtures.

\section{Discussion} \label{sec:discussion}

We now turn to discussing the implications of our interior structure simulation and gravity harmonics calculation results on UOP orbit design (Section \ref{sec:implication_UOP}), magnetic field generation in Uranus (Section \ref{sec:implication_magnetic}), and the condition of mixing in the interior of Uranus (Section \ref{sec:implication_mixing}).

\subsection{Implications for UOP Mission Design} \label{sec:implication_UOP}
Our gravity harmonics simulation results (Figure \ref{fig:jn_region}) imply that close-in polar orbits will be essential for UOP to detect higher order $J_n$ harmonics (in particular $J_6$ and $J_8$) beyond $J_4$. Higher order gravity harmonics, even if crude, are more binding on the $\rho(r)$ profile than precisely measured lower order harmonics \citep{movshovitz_promise_2022}. Therefore, we recommend that UOP orbital design should prioritize close passages and large inclinations.

Our claim that UOP could measure $J_6$ and possibly $J_8$ of Uranus, given favorable orbits, is supported by predictions for UOP gravity field measurement uncertainties \citep{Mazarico_Gravity_2023, parisi_uranus_2024}. In close-in polar orbits (solid blue line, Figure \ref{fig:jn_region}), UOP can confidently measure $J_6$, because predicted $J_6$ from \texttt{CORGI} models are at least $\sim2$ orders of magnitude greater than the predicted uncertainty. The predicted $J_8$ values of more than a half of all cases are greater than the predicted uncertainty, so $J_8$ is possibly detectable. However, the large relative uncertainty implies that a $J_8$ measurement by UOP will offer limited insights into the interior composition (Figure \ref{fig:j6j8_error_bar}). On the contrary, a UOP with equatorial orbits can only robustly measure $J_2$ and $J_4$, which have already been probed by V2, unless the interior density profile fortuitously produces unexpectedly high $J_6$ (dotted blue line, Figure \ref{fig:jn_region}).  Note that \cite{Mazarico_Gravity_2023} only predicted UOP uncertainties for $n=2$, 3, 4, but we log-linearly extrapolate the uncertainty to $J_6$ based on $J_2$ and $J_4$ uncertainties.

Juno and Cassini, which successfully constrained the high order gravity harmonics of Jupiter and Saturn, support the necessity of close-in polar orbits. During its Grand Finale, with closer-in and more highly inclined orbits than before the Grand Finale, the $J_n$ uncertainties of Cassini was improved by $\sim2$ orders of magnitude (black dotted and solid lines, Figure \ref{fig:jn_region}). Juno, on polar orbits, also has small uncertainties \citep{iess_measurement_2018}, even though the orbits are far from optimized for gravity science due to high eccentricity \citep{durante_juno_2022}.

The better $J_n$ precision offered by close-in polar orbits is due to the physical nature of the gravity field. Because a planet's gravity field is typically dominated by zonal harmonics with even $n$, polar orbits that probe a much wider latitude range provide higher sensitivity than equatorial orbits. Unnormalized $J_n$ harmonics directly measured by a spacecraft have a sensitive $r^{-n}$ dependence on distance from the planet. Given the same instrument sensitivity, a more close-in orbit is more sensitive to $J_n$, especially for higher degree $n$.

The prevalence of deep atmospheric winds in giant planets further supports the necessity to obtain high order $J_n$ harmonics. While our CMS model assumes uniform rotation, differential rotation is expected to be common in giant planets. Saturn's higher-than-predicted $J_6$--$J_{10}$ indicate differential rotation several thousands of kilometers deep into the interior \citep{iess_measurement_2019, militzer_models_2019}. Differential rotation produced by winds can increase high order $J_n$, even boosting them above the uncertainty limits, making $J_8$ and even higher order harmonics robustly measurable by UOP compared to the uniform rotation scenario. 

Constraints on deep atmospheric winds rely on high order $J_n$ values because lower order $J_n$ values probe the deep interiors, while higher order $J_n$ sample more of the deep atmospheric dynamics \citep[e.g.,][]{miguel_interior_2023}. Therefore, $J_2$ and $J_4$ are not sensitive to atmospheric density structure. Atmospheric winds, which can penetrate as deep as $\sim1/4\,R_p$ into giant planets \citep{militzer_models_2019}, may be overlooked if UOP measures only up to $J_4$. Uranus' atmosphere features strong zonal winds with velocities up to 200 m s$^{-1}$, so the rotation rate measured by radio signals and magnetic field by V2 may not represent its deep interior rotation rate \citep{helled_uranus_2010}. A combination of both lower and higher order $J_n$ measurements is therefore necessary to constrain both the deep interior and atmosphere rotation rates of Uranus, hence providing comprehensive insights into Uranus' interior.

\begin{figure*}
    \centering
    \includegraphics[width=0.85\linewidth]{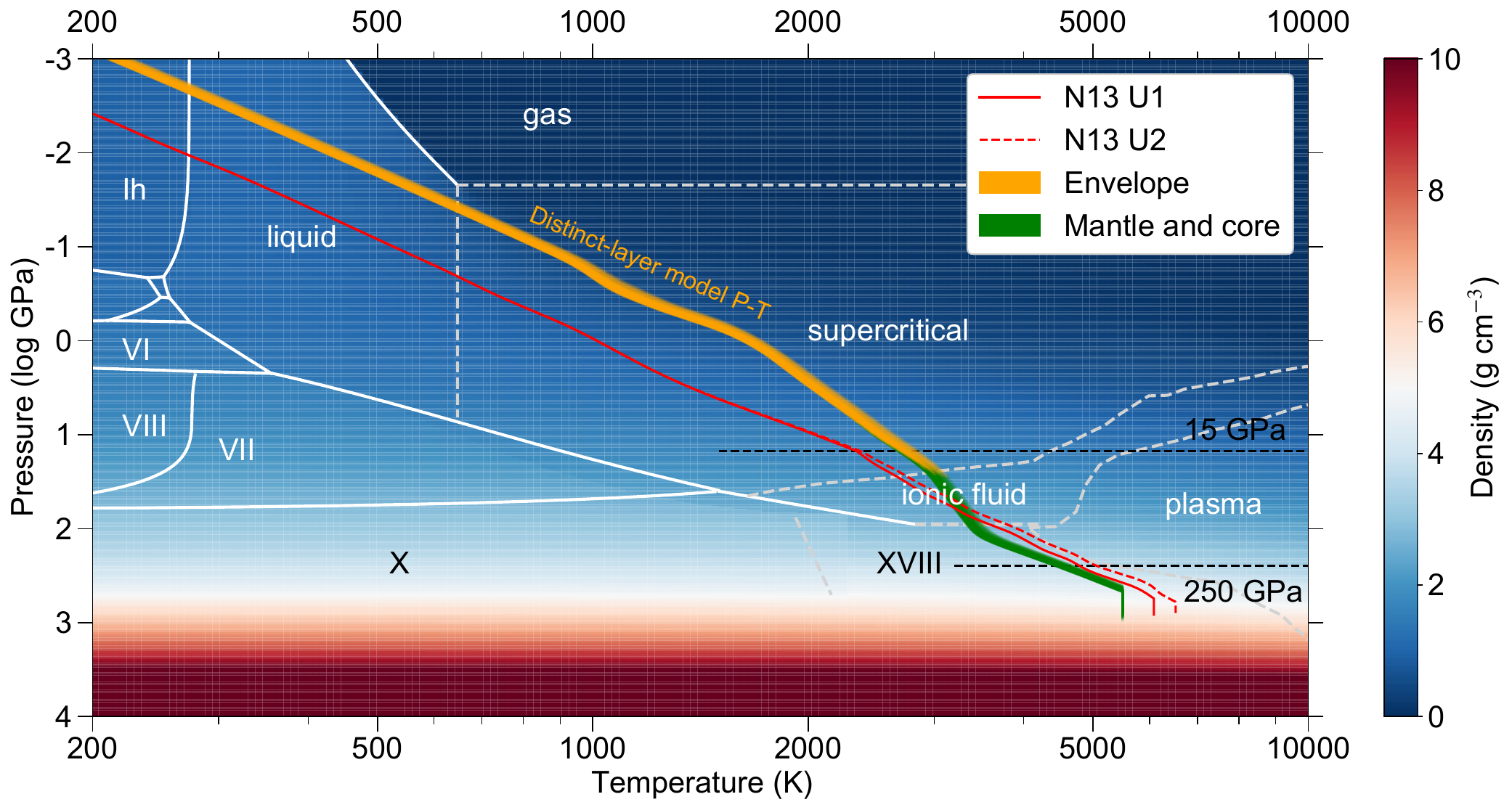}
    \caption{Uranus P-T profiles plotted over H$_2$O phase diagram. The shaded region show adiabatic P-T profiles of distinct-layer structure models, assuming mixed H$_2$O-H/He composition, in the (orange) envelope and (green) mantle and core. The red solid and red dashed lines show two Uranus P-T models from \cite{nettelmann_new_2013}. White lines are H$_2$O phase boundaries \citep{wagner_iapws_2002, dunaeva_phase_2010}. Background color represents H$_2$O density \citep{haldemann_aqua_2020}. Water phases are annotated. The region at which magnetic dynamo in Uranus is generated (15--250 GPa), according to the convective thin shell geometry \citep{stanley_convective-region_2004, stanley_numerical_2006}, is demarcated with black dashed lines. P-T profiles of distinct-layer Uranus models traverse superionic ice XVIII \citep{millot_nanosecond_2019}, ionic fluid, and supercritical phases, which have implications for magnetic field generation (see Section \ref{sec:implication_magnetic}) and interactions between the H/He envelope and the ice layer (see Section \ref{sec:implication_mixing}).}
    \label{fig:h2o_pt_aqua}
\end{figure*}

To self-consistently incorporate differential rotation into gravity models, one can approximate the wind profile as rotation on cylinders \citep{wisdom_differential_2016, militzer_models_2019}. In CMS model with differential rotation, instead of being a constant throughout the planet, rotation rate $\omega(l)$ becomes a function of distance from the rotation axis, $l$. Here we ignore differential rotation in order to efficiently explore various forward structure models. In future work, we plan to incorporate differential rotation into our CMS model and investigate the impact of winds on the predicted $J_n$ for Uranus.

Prior to the decadal survey, a planetary mission concept study (PMCS)\footnote{\href{https://smd-cms.nasa.gov/wp-content/uploads/2023/10/uranus-orbiter-and-probe.pdf}{https://smd-cms.nasa.gov/wp-content/uploads/2023/10/uranus-orbiter-and-probe.pdf}} was conducted for UOP. Results presented here echo some of the key science objectives presented in the PMCS. In the science traceability matrix presented in the PMCS, three key objectives require gravity field measurements to at least $J_8$, including (i) the bulk composition of Uranus, and the distribution with depth, (ii) whether Uranus has concentrated or diluted core, and how that is tied to its formation and tilt, and (iii) the deep interior rotation rate of Uranus, and whether its atmosphere is differentially rotating. Our results confirm that gravity harmonics up to $J_8$ is potentially achievable (Figure \ref{fig:jn_region}). Favorable orbital design will be pivotal for answering these key science questions.


\subsection{Implications for Uranian Dynamo} \label{sec:implication_magnetic}

We now switch to a discussion on the Uranian magnetic field, which provides a different set of constraints on the interior structure of Uranus compared to gravity field. The additional constraint offered by magnetic field is valuable, because as implied by our CMS results (Figure \ref{fig:j2j4_error_bar} and \ref{fig:j6j8_error_bar}), gravity harmonics measurements alone cannot resolve the distinct-layer and empirical structural degeneracy. As a result, the compositional degeneracy of Uranus remains unresolved, and the debate on whether Uranus has an ice-rich or rock-rich interior persists \citep[e.g.,][]{helled_interiors_2020, teanby_neptune_2020}.

The multipolar, non-axisymmetric magnetic field of Uranus can be produced by the convective thin shell dynamo geometry, which posits that the Uranian magnetic field is generated by convection in a shallow layer ($\sim0.5$--0.75 $R_p$) on top of a stably stratified fluid core \citep{stanley_convective-region_2004, stanley_numerical_2006}. Here, we assess the likelihood of ice-rich and rock-rich compositions by discussing whether they can generate a convective thin shell with high electrical conductivity at the predicted radius ($\sim0.5$--0.75 $R_p$, or $\sim15$--250 GPa).

\subsubsection{Ice-rich Interior Naturally Explains the Uranian Magnetic Field}
\texttt{CORGI} interior models suggest that the distinct-layer structure with an ice-rich composition is naturally consistent with the convective thin shell dynamo geometry. Ice-rich distinct-layer models are more physically consistent with the convective thin shell geometry because P-T profiles of distinct-layer models traverse regions where H$_2$O is both fluid and has high electrical conductivity (Figure \ref{fig:h2o_pt_aqua}). The dynamo-generating region between $\sim0.5$--0.75 $R_p$ is under pressures between $\sim15$--250 GPa according to our interior structure models. At such pressures, assuming adiabatic temperature profiles, the water-rich mantle of a distinct-layer planet traverses superionic ice XVIII, ionic fluid, finally reaching the supercritical phase where the H$_2$O layer is in contact with the H/He envelope at $\sim30$ GPa (Figure \ref{fig:h2o_pt_aqua}). The ionic fluid phase has a high electrical conductivity of $\gtrsim$ 3,000 S m$^{-1}$ \citep{french_diffusion_2010} while the superionic phase has a nearly metallic conductivity of $\sim10^4$ S m$^{-1}$, thereby accounting for the source of dynamo.

\begin{figure*}
    \centering
    \includegraphics[width=\linewidth]{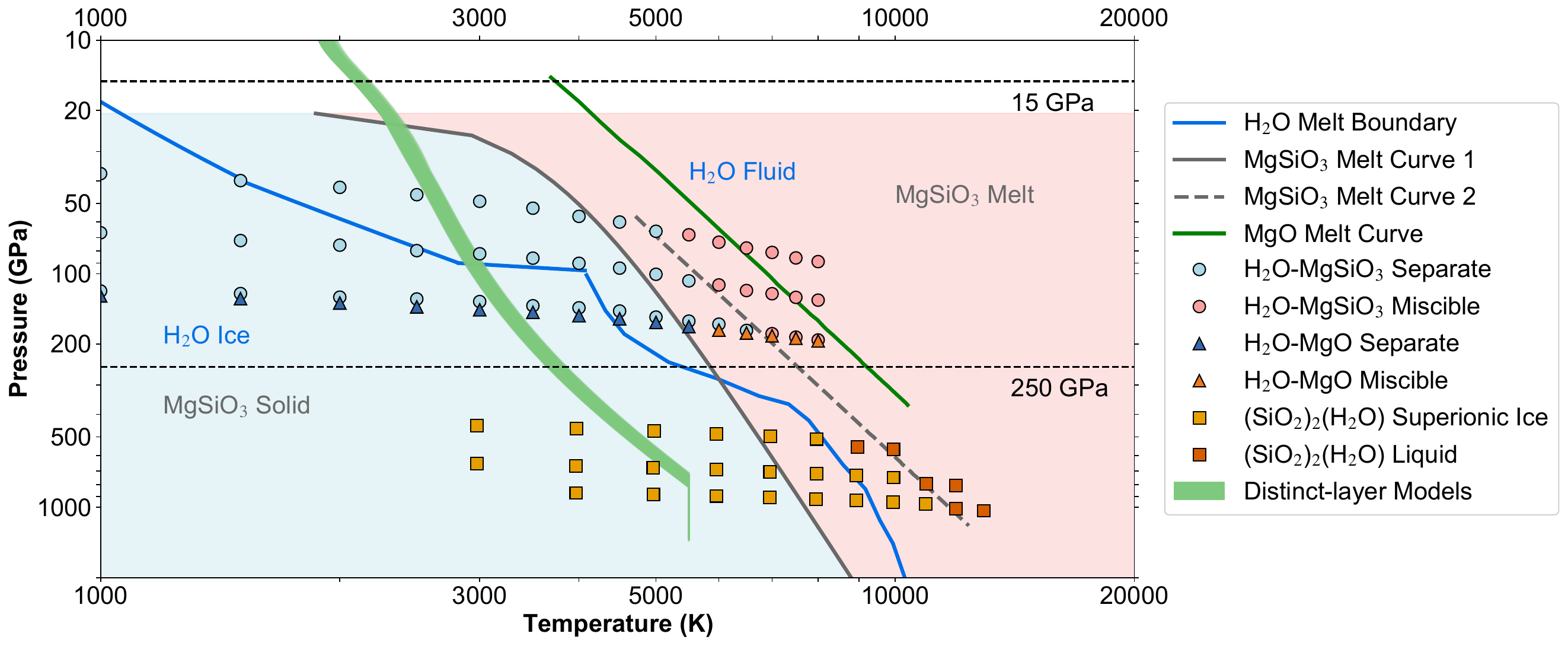}
    \caption{P-T phase diagram of MgSiO$_3$. Simulated data points for rock-water mixtures and H$_2$O melt curve (blue) are shown for context. The green shaded region outlines adiabatic P-T profiles of distinct-layer Uranus models. The gray solid line is the melt boundary of MgSiO$_3$ \citep{fratanduono_thermodynamic_2018}, separating MgSiO$_3$ solid (blue shaded region) from MgSiO$_3$ melt (red shaded region). The gray dashed line is MgSiO$_3$ melt curve measured by another group \citep{fei_melting_2021}. The green line shows the melt boundary of another rocky material, MgO \citep{soubiran_anharmonicity_2020}. Circles show DFT-MD simulations of the H$_2$O-MgSiO$_3$ system \citep{kovacevic_miscibility_2022} at P-T locations where they are separate (blue), or miscible (red). Triangles show DFT-MD simulations of the H$_2$O-MgO system \citep{kovacevic_homogeneous_2023} at locations where they are separate (blue), or miscible (red). Squares show simulations for the rock-water compound (SiO$_2$)$_2$(H$_2$O) \citep{gao_superionic_2022}, at locations where they are in a superionic ice phase (orange), or becomes liquid (red). In summary, rock-water mixtures become fluid and miscible at temperatures higher then predicted adiabatic P-T profiles of Uranus, implying that rock-rich interiors need to be hotter than ice-rich interiors to generate magnetic dynamo. The (SiO$_2$)$_2$(H$_2$O) compound may have implications for the deep interior magnetic field (see Section \ref{sec:rg_magnetic_field}).}
    \label{fig:silicate_phase_diagram}
\end{figure*}

However, there is a caveat that the physical properties of SPI layers or H$_2$O-H/He mixture layers may differ from those of pure H$_2$O. The addition of CH$_4$, NH$_3$, and H/He can modify locations of phase boundaries and change electrical conductivities relative to pure H$_2$O conductivity. Therefore, laboratory experiments that have been performed on pure H$_2$O should also be performed on icy mixtures to allow confident determination of the source of the Uranian magnetic field. Recently, \cite{militzer_phase_2024} demonstrated using ab initio simulations that phase separation of H$_2$O-CH$_4$-NH$_3$ mixture can reproduce the convective thin shell geometry suggested by \cite{stanley_convective-region_2004, stanley_numerical_2006}.

\subsubsection{\texttt{CORGI} Results Are Agnostic about whether Rock-rich Interior Can Explain the Uranian Magnetic Field} \label{sec:rg_magnetic_field}
Whether a Uranus model with a rock-rich interior can generate the observed magnetic dynamo remains unknown. As opposed to ice-rich interior models, where EOS and electrical conductivity of H$_2$O are better studied, physical properties of rock-rich mixtures under the P-T conditions relevant for the interior of Uranus are less well-constrained. Here, we outline two physical properties that, if clarified by future experiment and ab initio simulation, will inform us about dynamo generation in a rock-rich Uranus interior model.

The first relevant physical property is rock-ice miscibility. Rock-rich interior models generally assume a smooth compositional gradient that implies extensive mixing between rock and ice. To generate a dynamo, rock and ice must be miscible and convective between $\sim15$--250 GPa. DFT-MD simulations of the H$_2$O-MgSiO$_3$ system \citep{kovacevic_miscibility_2022} show that H$_2$O and MgSiO$_3$ become miscible when temperature exceeds the melting temperature of MgSiO$_3$. Experimentally determined MgSiO$_3$ melt curves are much hotter than the predicted adiabatic P-T profiles of Uranus (Figure \ref{fig:silicate_phase_diagram}). Therefore, to generate the observed magnetic field, rock-rich interior models must have hotter P-T profiles than distinct-layer models. Such hot P-T profiles may be produced by thermal boundary layers or superadiabatic temperature gradient, which are both plausible \citep[e.g.,][]{nettelmann_uranus_2016, podolak_effect_2019}.
The second relevant physical property is electrical conductivity. High electrical conductivity is required for dynamo generation, but constraints on the electrical conductivities of rocky materials and rock-ice mixtures under the relevant ($\sim15$--250 GPa) pressures are lacking. Existing simulations are generally performed assuming purely rocky composition (e.g., pure MgO, SiO$_2$, and MgSiO$_3$) and cover pressures and temperatures outside of the parameter space of interest ($\gtrsim500$ GPa and $\gtrsim10,000$ K, see e.g., \citealt{soubiran_electrical_2018} and \citealt{guarguaglini_electrical_2021}).
Therefore, we call for experiments and simulations that can derive electrical conductivities of rock-ice mixtures at pressures and temperatures relevant for the dynamo generating region within Uranus (Figure \ref{fig:silicate_phase_diagram}). \cite{gao_superionic_2022} presented a simulation revealing that the rock-water compound (SiO$_2$)$_2$(H$_2$O) and rock-hydrogen compound SiO$_2$H$_2$ can exist under pressures $>450$ GPa and $>650$ GPa, respectively. These compounds even exhibit superionic behaviors under high pressures and temperatures (Figure \ref{fig:silicate_phase_diagram}). While such extreme pressures are more relevant to the conditions in Uranus' core ($<0.3\,R_p$) than to the convective thin shell, they offer valuable constraints on the electrical conductivity of a mixed material that is possibly present inside a rock-rich Uranus interior.

Future computational and experimental efforts into investigating the physical properties of rock-ice mixtures will be critical to understand the Uranian magnetic field, even though such investigations are much more challenging than studying single-component systems.

\subsection{Implications for Mixing within Uranus} \label{sec:implication_mixing}
Our interior models imply that mixing -- among ices (H$_2$O, CH$_4$, and NH$_3$), between ices and H/He, and between ices and rock -- are prevalent in intermediate-sized planets. In particular, H/He-rich atmosphere layers in our models are in direct contact with supercritical water (Figure \ref{fig:h2o_pt_aqua}), an excellent solvent, implying H/He mixing with H$_2$O. The likelihood of such mixing is also supported by the consistency between mixed H$_2$O-H/He models and measured $J_n$ harmonics.

Whether hydrogen and water are miscible under the P-T conditions relevant for the interior of Uranus, however, is still under debate. Experimental data obtained for an impure system containing hydrogen, water, and silicate, which is representative of Uranus' interior especially if the core is diffuse, suggest that H$_2$ and H$_2$O are immiscible \citep{bali_water_2013}. On the contrary, ab initio simulations in the range of 2--70 GPa and 1000--6000 K, intersecting our distinct-layer adiabatic P-T profiles and partially encompassing the supercritical, ionic fluid, and plasma phases, predicted that H$_2$ and H$_2$O are fully miscible \citep{soubiran_miscibility_2015}. \cite{bailey_thermodynamically_2021} presented thermodynamically self-consistent models of the interiors of Uranus and Neptune under the assumption of H$_2$-H$_2$O immiscibility. Interestingly, they concluded that $Z_{\rm atm}$ of Uranus should be very small, on the order of $\lesssim0.01$, which disagrees with our result that large $Z_{\rm atm}\sim0.25$ is favored to produce $J_n$ consistent with V2 measurements. This discrepancy arises because, by allowing $Z$ to take arbitrary values, we are essentially assuming that H$_2$ and H$_2$O are miscible by any fraction, while \cite{bailey_thermodynamically_2021} assumed immiscibility. Because experimental data and ab initio simulations on H$_2$-H$_2$O miscibility disagree \citep{bali_water_2013, soubiran_miscibility_2015}, further investigations are needed to inform us about which fundamental assumption is correct.

A thick supercritical water ocean can potentially lead to atmospheric signatures detectable by remote reconnaissance. An extensive supercritical water ocean would serve as a chemical sink and source for the atmosphere. 
Supercritical water is an excellent solvent for both polar and nonpolar compounds \citep{weingartner_supercritical_2005}. Spectrally prominent atmospheric species such as CH$_4$ and NH$_3$ may be highly soluble in supercritical water, which can limit the amount of upward transport of these molecules. On the contrary, if the water layer is replaced by a more realistic SPI mixture rich in C, N and O \citep[e.g.,][]{bethkenhagen_planetary_2017, guarguaglini_laser-driven_2019}, it may serve as a reservoir supplying these atoms to the atmosphere, replenishing losses due to photochemistry and atmospheric escape to space. In either case, atmospheric features accessible to remote observers in a planet with supercritical ocean may differ dramatically from a planet without. For further discussions on interior-atmosphere interactions on intermediate-sized planets, see \cite{yu_how_2021} and \cite{hu_unveiling_2021}.

Our results motivate experimental study of miscibility and diffusivity in a mixture system involving H, C, N, and O under high pressures (tens of GPa) and temperatures ($\sim2000$--5000 K), in order to more precisely constrain the interactions between H/He envelope and supercritical water.


\section{Conclusions} \label{sec:conclusion}
Uranus, the target of next Flagship mission proposed by the Planetary Science and Astrobiology Decadal Survey 2023--2032 \citep{decadal_survey_2023}, has an ambiguous interior structure that requires the combination of multiple observables to resolve. Among these observables, gravity field measurements offer some of the most direct insights into the density distribution within the deep interior and deep atmosphere of Uranus, while magnetic field provides additional indirect constraints on conductivity and fluid flows. In this paper, we simulate the interior structure and zonal gravity harmonics of Uranus with newly developed \texttt{CORGI}, a code package with forward modeling, inverse retrieval, and concentric Maclaurin spheroid gravity harmonics modules. We simulate two common classes of interior models for Uranus: fully differentiated distinct-layer models with adiabatic ice-rich mantles, and empirical density models that allow smooth transitions and compositional gradients (Figure \ref{fig:RGIG_rhoz_baseline}). For the distinct-layer models, layer mass fractions are obtained by running the interior retrieval model (Figure \ref{fig:uranus_retrieval}), while the empirical $\rho(z)$ profiles are obtained via parametrization (Section \ref{sec:empirical_density_model}; see also \citealt{movshovitz_promise_2022}). Then, we simulate the zonal gravity harmonics of all the forward models up to $J_{30}$ with numerical precision $\sim10^{-12}$ using the CMS method \citep{hubbard_concentric_2013}. We further discuss whether ice-rich and rock-rich interior compositions are consistent with the multipolar and non-axisymmetric magnetic field of Uranus.

Major implications of our interior and gravity harmonics modeling results include the following.
\begin{enumerate}
    \item High degrees of mixing are required for interior models of Uranus to be consistent with $J_2$ and $J_4$ measured by V2. Uranus either has a smooth density profile as suggested by empirical models, which naturally requires extensive mixing, or if it has a distinct-layer structure, heavy element mass fraction, $Z$, must low in the mantle or high in the atmosphere (Figure \ref{fig:j2j4_error_bar}). Quantitatively, if Uranus' mantle and atmosphere are composed of H$_2$O-H/He mixture, the retrieved $Z_{\rm mantle}$ and $Z_{\rm atm}$ are roughly 0.85 and 0.28, respectively, in agreement with previous results \citep[e.g.,][]{nettelmann_new_2013}. If the mantle of Uranus consists of synthetic planetary ice (H$_2$O-CH$_4$-NH$_3$ at solar C:N:O elemental ratio), a low $Z_{\rm mantle}$ of 0.85 or a high $Z_{\rm atm}$ of 0.25 is required for the predicted $J_2$ and $J_4$ to be consistent with V2 measurements.
    \item UOP $J_2$ and $J_4$ measurements will be able to distinguish between high and low atmospheric metallicity models, while Voyager 2 data suffer from large uncertainties in $J_4$ and in data analysis (Figure \ref{fig:j2j4_error_bar}b and c).
    \item $J_6$ measurement by the UOP can potentially rule out some uniformly rotating SPI models. $J_8$ measurements by the UOP will offer limited constraints on the interior given current uncertainty predictions (Figure \ref{fig:j6j8_error_bar}b and c).
\end{enumerate}

Our results further suggest that a pure H$_2$O mantle, which is a common simplification for modeling the interiors of exoplanets, is totally inconsistent with measurements. To realistically simulate extrasolar intermediate-sized planets, the community should move away from the oversimplified pure H$_2$O ice assumption and adopt a mixed-composition (H$_2$O-CH$_4$-NH$_3$) ice with some light elements (H/He) instead. \texttt{CORGI} offers helpful tools for simulating such mixed-composition interiors using EOS based on linear mixing approximation.

In terms of orbit design, our CMS results suggest that close-in polar orbits are necessary for the UOP mission to measure higher order harmonics beyond $J_2$ and $J_4$, which are already measured by V2 (Figure \ref{fig:jn_region}). Given the UOP uncertainties predicted by \cite{parisi_uranus_2024}, $J_6$ can be robustly measured. $J_8$ can potentially be constrained, albeit with large uncertainties that make it less useful to constrain interior structure than $J_6$ (Figure \ref{fig:j6j8_error_bar}). These high order harmonics are helpful for narrowing the parameter space of $\rho(r)$ profiles of Uranus \citep{movshovitz_promise_2022}, potentially reducing the structural degeneracy between distinct-layer and empirical models \citep[e.g.,][]{helled_interiors_2020, teanby_neptune_2020}. In addition, $J_6$ and $J_8$ probe depths that $J_2$ and $J_4$ are not sensitive to and may reveal deep atmospheric winds in Uranus.

Our interior model results hint that ice-rich distinct-layer interior can naturally explain the multipolar, non-axisymmetric magnetic field of Uranus due to its consistency with the convective thin shell dynamo geometry \citep{stanley_convective-region_2004, stanley_numerical_2006}. This is because P-T profiles of distinct-layer models traverse a region where H$_2$O is both fluid and has high electrical conductivity (Figure \ref{fig:h2o_pt_aqua}). \texttt{CORGI} results remain agnostic to whether rock-rich interior composition is consistent with the convective thin shell dynamo geometry, because physical properties of rock-ice mixtures remain underexplored (Figure \ref{fig:silicate_phase_diagram}). We therefore call for experiment and ab initio simulation for rock-ice mixtures under the relevant pressure and temperature conditions.

Finally, our results imply that mixing between H/He-dominated envelope and the volatile-rich ice layer may be common in intermediate-sized planets. The atmosphere of Uranus is in direct contact with supercritical H$_2$O (Figure \ref{fig:h2o_pt_aqua}). Supercritical water is an excellent solvent for H$_2$, CH$_4$, and NH$_3$, which may produce compositional gradient at the atmosphere-ice layer boundary and can even alter atmospheric signatures produced by CH$_4$ and NH$_3$, which are accessible to remote spectroscopic observations.
The interior composition of intermediate-sized planets, including Uranus, Neptune, and extrasolar sub-Neptunes and Neptune-like planets, remains a mystery despite such planets are ubiquitous.
Constraining the interior composition of exoplanets is highly challenging. For exoplanets, mass and radius, often with large error bars, are the only currently accessible observables offering constraints on bulk density.
UOP will offer a rare and highly valuable opportunity to study an intermediate-sized planet in situ. Gravity field measurements by UOP will offer us some of the most robust constraints on the interior mass distribution of Uranus, potentially ruling out some plausible interior models and reducing the compositional and structural degeneracy. Therefore, we recommend a close-in, polar orbit for UOP to best leverage this opportunity and increase the scientific yield of gravity science. Gravity science alone, however, is insufficient to fully resolve the compositional and structural degeneracy of Uranus. Additional constraints are therefore needed.
Magnetic field offers one of such additional constraints: the measured magnetic field of Uranus requires a shallow convective and highly electrically conductive layer. To accurately model the source of magnetic dynamo, however, requires new knowledge about physical properties including rock-ice miscibility and electrical conductivities of mixed-composition ices and rock-ice mixtures. 
Insights into the interior composition of Uranus will help to reduce the compositional degeneracy of intermediate-sized exoplanets in combination with remote atmospheric reconnaissance. \\
\\
\noindent
We thank the anonymous referees for constructive comments that significantly improved the quality of this manuscript. We thank Jack Wisdom for helpful discussions on CMS convergence. Z.L. acknowledges funding from the Center for Matter at Atomic Pressures (CMAP), a National Science Foundation (NSF) Physics Frontiers Center, under award PHY-2020249. The authors acknowledge the MIT SuperCloud and Lincoln Laboratory Supercomputing Center for providing high performance computing resources that have contributed to the research results reported within this paper.

%



\software{\texttt{Matplotlib} \citep{Hunter_2007_matplotlib}, \texttt{NumPy} \citep{harris2020array}, \texttt{SciPy} \citep{2020SciPy-NMeth}, \texttt{emcee} \citep{foreman-mackey_emcee_2013, foreman-mackey_emcee_2019}.}



\appendix

\section{Detailed Results and Corner Plot for Interior Retrieval} \label{sec:appendix_corner_plot}
\begin{figure}[t!]
    \centering
    \includegraphics[width=0.95\linewidth]{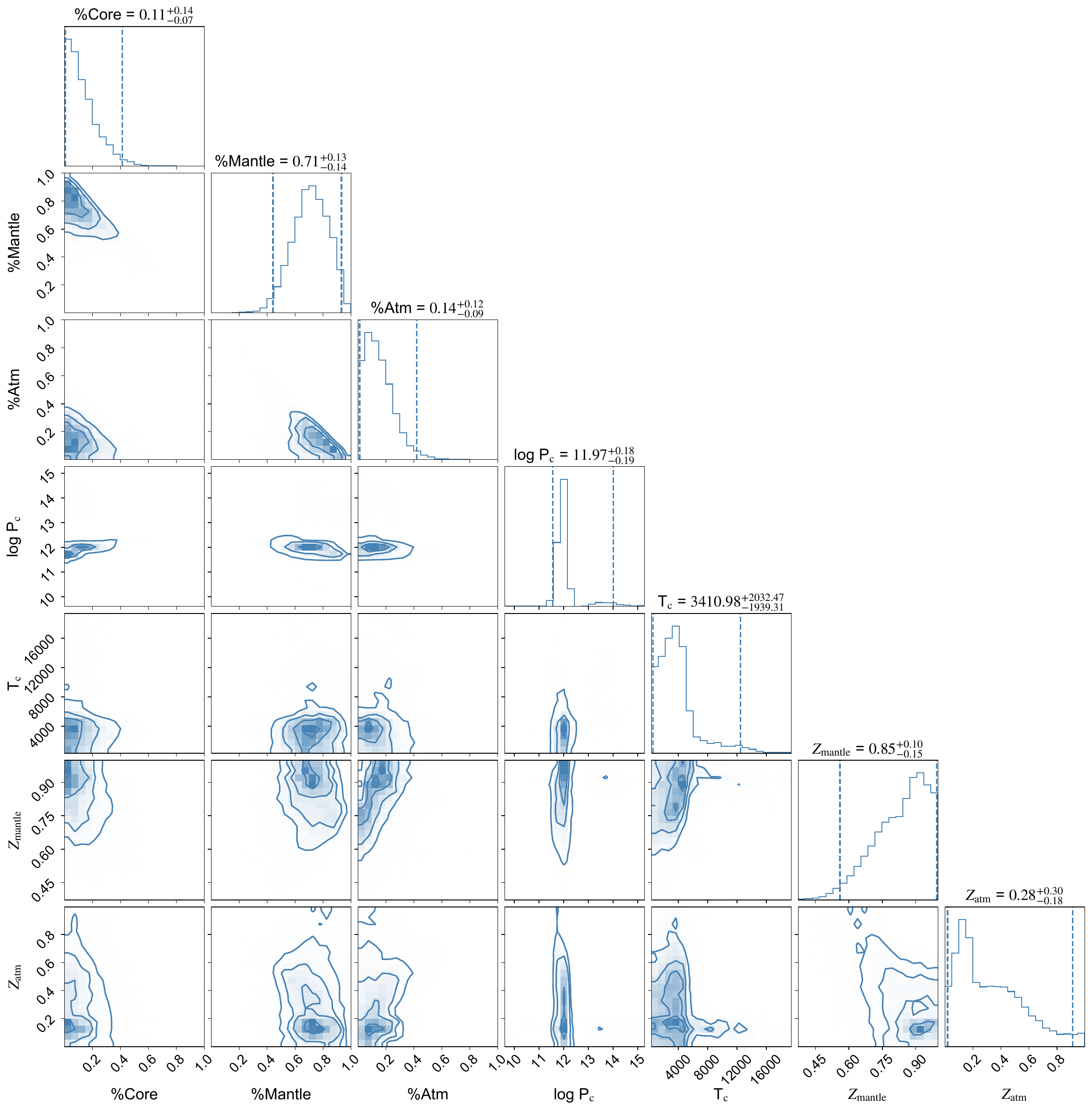}
    \caption{Corner plot showing the retrieved core mass fraction (\%Core), mantle mass fraction (\%Mantle), atmosphere mass fraction (\%Atm), central pressure ($\log_{10} P_{\rm c}$, in GPa), central temperature ($T_{\rm c}$, in K), heavy element mass fraction in the mantle ($Z_{\rm mantle}$), and heavy element mass fraction in the atmosphere ($Z_{\rm atm}$) of Uranus. Here, the mixed H$_2$O-H/He composition is assumed (see Section \ref{sec:inverse_model}). The dashed lines in the histograms outline $3\,\sigma$ ranges, while the $\pm$ values in titles are $1\,\sigma$ ranges. Interior retrieval predicts a small core, massive mantle, and significant mixing of heavy elements into the atmosphere and mixing of light elements into the mantle.}
    \label{fig:corner_plot}
\end{figure}

As mentioned in Section \ref{sec:inverse_model}, \texttt{CORGI} has a planet interior retrieval module that retrieves the most likely layer mass fractions, central pressure, central temperature, and $Z$ values (if applicable) for a planet. In the main text, we use the mixed H$_2$O-H/He composition as an example to demonstrate the retrieval module. The corner plot for this composition, which shows best-fit core mass fraction (\%Core), mantle mass fraction (\%Mantle), atmosphere mass fraction (\%Atm), central pressure ($\log_{10} P_{\rm c}$, where $P_c$ is in Pa), central temperature ($T_{\rm c}$), heavy element mass fraction in the mantle ($Z_{\rm mantle}$), and heavy element mass fraction in the atmosphere ($Z_{\rm atm}$), is shown in Figure \ref{fig:corner_plot}.

Retrieval results for other compositional models are summarized below.
\begin{itemize}
    \item Pure H$_2$O mantle: $x_{\rm core} = 0.09^{+0.10}_{-0.06}$, $x_{\rm mantle} = 0.75^{+0.09}_{-0.10}$, $x_{\rm atm} = 0.14^{+0.06}_{-0.04}$, $\log_{10} P_c = 12.07^{+0.11}_{-0.12}$, and $T_c = 3971.16^{+2405.01}_{-2591.82}$ K.
    \item SPI, $Z_{\rm mantle}=85\%$, $Z_{\rm atm}=12.5\%$: $x_{\rm core} = 0.10^{+0.09}_{-0.07}$, $x_{\rm mantle} = 0.77^{+0.09}_{-0.10}$, $x_{\rm atm} = 0.11^{+0.07}_{-0.04}$, $\log_{10} P_c = 11.98^{+0.14}_{-0.16}$, and $T_c = 4022.89^{+2553.03}_{-2346.84}$ K.
    \item SPI, $Z_{\rm mantle}=85\%$, $Z_{\rm atm}=25\%$: $x_{\rm core} = 0.10^{+0.08}_{-0.06}$, $x_{\rm mantle} = 0.75^{+0.09}_{-0.10}$, $x_{\rm atm} = 0.14^{+0.07}_{-0.04}$, $\log_{10} P_c = 11.97^{+0.12}_{-0.15}$, and $T_c = 3937.07^{+1746.67}_{-2239.28}$ K.
    \item SPI, $Z_{\rm mantle}=95\%$, $Z_{\rm atm}=12.5\%$: $x_{\rm core} = 0.09^{+0.08}_{-0.06}$, $x_{\rm mantle} = 0.78^{+0.08}_{-0.10}$, $x_{\rm atm} = 0.11^{+0.08}_{-0.04}$, $\log_{10} P_c = 11.98^{+0.16}_{-0.14}$, and $T_c = 4301.38^{+3008.76}_{-2444.51}$ K.
    \item SPI, $Z_{\rm mantle}=95\%$, $Z_{\rm atm}=25\%$: $x_{\rm core} = 0.10^{+0.09}_{-0.07}$, $x_{\rm mantle} = 0.75^{+0.09}_{-0.09}$, $x_{\rm atm} = 0.14^{+0.07}_{-0.04}$, $\log_{10} P_c = 11.98^{+0.13}_{-0.17}$, and $T_c = 4099.66^{+2146.16}_{-2306.99}$ K.
    \item SPI, $Z_{\rm mantle}=95\%$, $Z_{\rm atm}=12.5\%$: $x_{\rm core} = 0.10^{+0.07}_{-0.06}$, $x_{\rm mantle} = 0.76^{+0.08}_{-0.09}$, $x_{\rm atm} = 0.12^{+0.07}_{-0.04}$, $\log_{10} P_c = 11.96^{+0.13}_{-0.18}$, and $T_c = 4369.66^{+3488.73}_{-2727.03}$ K.
    \item SPI, $Z_{\rm mantle}=95\%$, $Z_{\rm atm}=25\%$: $x_{\rm core} = 0.10^{+0.08}_{-0.06}$, $x_{\rm mantle} = 0.74^{+0.09}_{-0.09}$, $x_{\rm atm} = 0.14^{+0.07}_{-0.05}$, $\log_{10} P_c = 11.96^{+0.13}_{-0.16}$, and $T_c = 4143.05^{+3426.98}_{-2540.17}$ K.
\end{itemize}

\section{Detailed Description of the CMS Method} \label{sec:CMS_details}

In Section \ref{sec:CMS_model_details}, we present a qualitative overview of the CMS method. Here, we describe the formalism of the CMS method in detail.

The first step is ensuring consistency of units. Because the forward structure model outputs are in SI units, which involves numbers spanning many orders of magnitudes, it is convenient to express pressure, density, and potential in dimensionless planetary units (pu) as
\begin{equation}
\begin{split}
    P_{\rm pu} &\equiv \left(\frac{R_{\rm eq}^4}{GM_p^2}\right) P_{\rm SI}, \\
    \rho_{\rm pu} &\equiv \left(\frac{R_{\rm eq}^3}{M_p}\right) \rho_{\rm SI}, \\
    U_{\rm pu} &\equiv \left(\frac{R_{\rm eq}}{GM_p}\right) U_{\rm SI},
\end{split}
\end{equation}
where $R_{\rm eq}$ is equatorial radius in meters. Note that $M_p$ here is planet mass in unit of kg and is not to be confused with the dimensionless mass $M$ to be discussed in Section \ref{sec:grav_potential}. Unless otherwise noted, variables defined below are in planetary units.

Now we describe the CMS method in detail. We use $N_L=128$ layers following \cite{hubbard_concentric_2013}. Even though numerical precision of $J_n$ predictions improve as $N_L$ increases, the computation time also drastically increase without acceleration \citep[e.g.,][]{militzer_models_2019}. To rapidly cover a wide parameter space, we stick to $N_L=128$ layers. On each layer, there are $N_M$ Gaussian quadrature points with angles $\mu_m = \cos(\theta_m)$ and Gaussian quadrature weights $w_m$. For simulation of even $J_n$ harmonics, Gaussian quadrature guarantees sufficient numerical precision as long as $N_M > n_{\rm max}$, the maximum order modelled. Here we assume $N_M = 48$ following \cite{hubbard_concentric_2013}, which yields a numerical precision of $\sim 10^{-12}$ or the floating point precision of the computer, whichever one is lower. The position of each grid point is defined by two parameters, namely its distance from the center $r_{im}$, and the cosine of its angle from the equator $\mu_m$, where $\mu = 1$ at the north pole, $\mu = 0$ on the equator, and $\mu = -1$ at the south pole. The model assumes north-south symmetry, implying that $r_i(\mu_m) = r_i(-\mu_m)$. A normalized shape function is introduced for convenience
\begin{equation}
    \zeta_i(\mu_m) \equiv \zeta_{im} \equiv \frac{r_i(\mu_m)}{r_i(0)} \leq 1,
\end{equation}
where $r_i(0)$, the equatorial radius at the $i$th layer, is kept constant throughout iteration. At the surface, $r_0(0) = R_{\rm eq}$ is the equatorial radius of the planet. Initially, we assume a perfectly spherical planet with $\zeta = 1$ everywhere. Two more parameters are defined for convenience, namely the ratio of the equatorial radius of the $i$th layer to the planet's equatorial radius
\begin{equation}
    \lambda_i \equiv \frac{r_i(0)}{r_0(0)},
\end{equation}
and the density difference between two adjacent spheroids
\begin{equation}
    \delta_i = \begin{cases}
        \rho_i - \rho_{i-1}, & {i>0}, \\
        \rho_0, & {i=0}.
    \end{cases}
\end{equation}
Note that $\lambda_i$ is fixed, because all $r_i(0)$ are fixed throughout integration.

We may now define a potential function such that the total potential of a grid point with index $(i, m)$ is expressed as $U_{im} \equiv U_i(\zeta_{im}, \mu_m)$. By definition, a CMS run is converged when $U_i$ is the same for every $\mu_m$, on each layer surface. Before convergence, $U_{im}$ deviates from a reference value. This deviation is expressed as 
\begin{equation}
    f_{im}(\zeta_{im}, \mu_m) = U_i(\zeta_{im}, \mu_m) - U_i(1, 0),
\end{equation}
where $U_i(1, 0)$, or $U_i$ for short, is the potential of a reference point on the equator of the $i$th layer surface. The gravitational part of $U_i$ is derived in Section \ref{sec:grav_potential} and the centrifugal part in Section \ref{sec:centri_potential}. The goal of CMS is therefore to minimize $f_{im}$ until it falls below some small tolerance. This is achieved by updating the shape function using Newton steps
\begin{equation}
    \zeta_{im}^\text{(new)} = \zeta_{im} - \frac{f_{im}(\zeta_{im})}{f'_{im}(\zeta_{im})},
\end{equation}
where $f'_{im}(\zeta_{im}) = df_{im}(\zeta_{im}) / d\zeta_{im}$ is the derivative. The analytical expression for $f'_{im}(\zeta_{im})$ is derived in Appendix \ref{sec:appendix_newton_step}.

Once the shape functions are updated, we may calculate the new $J_n$ and hence the new potential $U_i$. The pressure at each grid point can then be updated with the hydrostatic equilibrium assumption ($\nabla P = \rho \nabla U$) as
\begin{equation}
    P_{i}^{\rm (new)} = P_{i-1}^{\rm (new)} + \rho_{i-1}(U_i - U_{i-1}),
\end{equation}
starting from fixed surface pressure $P_0$. \cite{militzer_models_2019} used a fixed $P_0$ of 0.1 bar. Here we use $P_0 = P_{\rm surf}$, where $P_{\rm surf}$ is the surface pressure calculated by the forward planet structure model. Density at each grid point can be updated as
\begin{equation}
    \rho_i^{\rm (new)} = \rho(\bar{P}),
\end{equation}
where $\bar{P} = \frac{1}{2}(P_{i+1}^{\rm (new)} + P_i^{\rm (new)})$ is the mean pressure and $\rho(P)$ comes from the EOS, assuming $T$ at a layer surface remains constant. To avoid repeatedly calling physical EOS databases, which involves large tables and slows down the computation, we define an equivalent EOS for each planet model by simply interpolating the $\rho(P)$ profile calculated by the forward structure model. 

\subsection{Gravitational Potential} \label{sec:grav_potential}
The total potential $U_{im} = V_{im} + Q_{im}$ at each grid point has both gravitational and centrifugal components. Here, we give the expressions for gravitational potential. For simplicity, we include only the equations relevant for implementing the CMS method without derivation (see \citealt{hubbard_concentric_2013} for details).

Zonal gravity harmonics $J_n$ are given by
\begin{equation} \label{eq:zonal_Jn}
    J_{n} = -\frac{2\pi}{M_p R_{\rm eq}^{n}} \int_{-1}^{+1} d\mu \int_0^{r_{\rm max}(\mu)} dr\,r^{n+2} P_n(\mu) \rho(r, \mu),
\end{equation}
where $P_n(\mu)$ is the Legendre polynomial. $J_0$ integrates over the entire planet mass and is conventionally normalized to $-1$. The gravity harmonics of a grid point inside the planet have both interior ($J_{i, n}$) and exterior ($J'_{i, n}$ and $J''_{i, n}$) components. To avoid repeatedly multiplying and dividing by large factors, \cite{hubbard_concentric_2013} normalized the interior and exterior harmonics as
\begin{equation}
\begin{split}
    \widetilde{J}_{i,n} &\equiv \frac{J_{i,n}}{\lambda_i^n}, \\
    \widetilde{J}_{i,n}' &\equiv J_{i,n}' \lambda_i^{(n+1)}.
\end{split}
\end{equation}
Then, \cite{hubbard_concentric_2013} derived the expressions for interior gravity harmonics as
\begin{equation}
    \widetilde{J}_{i,n} = -\frac{1}{n+3}\frac{2\pi}{M} \delta_i \lambda_i^3 \int_{-1}^{+1} d\mu P_n(\mu) \zeta_i(\mu)^{n+3},
\end{equation}
and the exterior gravity harmonics as
\begin{equation}
    \widetilde{J}_{i,n}' = -\frac{1}{2-n}\frac{2\pi}{M} \delta_i \lambda_i^3 \int_{-1}^{+1} d\mu P_n(\mu) \zeta_i(\mu)^{2-n},
\end{equation}
with a special case when $n=2$
\begin{equation}
    \widetilde{J}_{i,n}' = -\frac{2\pi}{M} \delta_i \lambda_i^3 \int_{-1}^{+1} d\mu P_n(\mu) \log(\zeta_i),
\end{equation}
and
\begin{equation}
    \widetilde{J}_{i,0}'' = \frac{2\pi \delta_i R_{\rm eq}^3}{3M}.
\end{equation}
Note that $\widetilde{J}_{i,n}$, $\widetilde{J}_{i,n}'$, and $\widetilde{J}_{i,0}''$ are dimensionless. $M$ in the above equations is dimensionless total mass of the planet, given by
\begin{equation}
    M = \frac{2\pi}{3} \sum_{i=0}^{N_L-1} \delta_i \lambda_i^3 \int_{-1}^{+1} d\mu\,\zeta_i(\mu)^3.
\end{equation}
In practice, the integrals over $\mu$ above are approximated by Gaussian quadrature sums as
\begin{equation}
    \int_{-1}^{+1} d\mu\ f(\mu) \approx \sum_{m=1}^{N_M} w_m f(\mu_m).
\end{equation}
Given $\widetilde{J}_{i,n}$, $\widetilde{J}_{i,n}'$, and $\widetilde{J}_{i,n}''$, the gravitational potential for a point on the $i$th layer surface ($i \neq 0$) can be expressed as
\begin{equation} \label{eq:Vi_definition}
    V_i(\zeta_i, \mu) = -\frac{1}{\zeta_i \lambda_i} 
    \left[
    \sum_{j=i}^{N_L-1} \sum_{n=0}^{\infty} \widetilde{J}_{j,n}
        \left( 
        \frac{\lambda_j}{\lambda_i \zeta_i}
        \right)^n
        P_n(\mu) \right. + \left.\sum_{j=0}^{i-1} \sum_{n=0}^\infty \widetilde{J}_{j,n}'
        \left( 
        \frac{\lambda_j \zeta_i}{\lambda_j}
        \right)^{n+1}
        P_n(\mu) + \sum_{j=0}^{i-1} \widetilde{J}_{j,0}'' \lambda_i^3 \zeta_i^3
    \right].
\end{equation}
On the equator of the surface layer, $V$ is given by
\begin{equation}
    V_{i=0}(1,0) = - \sum_{n=0}^\infty P_n(0) J_n,
\end{equation}
where $J_n$ are the standard gravity harmonics defined in Equation (\ref{eq:zonal_Jn}) and can be expressed as
\begin{equation}
    J_n = \sum_{i=0}^{N_L-1} \lambda_i^n \widetilde{J}_{i,n}.
\end{equation}
In practice, CMS results converge rapidly with increasing degree $n$, and the upper bound of the summation over $n$ is replaced by some finite maximum degree $n_{\rm max}$. Our model adopts $n_{\rm max}=30$ by default.

\subsection{Centrifugal Potential} \label{sec:centri_potential}
When assuming uniform rotation with rate $\omega$, the centrifugal potential is simply
\begin{equation} 
    Q(l) = \frac{1}{2}l^2 \omega^2 
\end{equation}
where $l = r \sin(\theta)$ is the distance from the rotation axis. Noting that $\sin^2(\theta) = 1 - \mu^2$ and $r_i(\mu_m) = \zeta_{im} r_i(0)$, $Q$ can be rewritten using the shape function as
\begin{equation} \label{eq:Qi_deifinition}
    Q_i(\zeta_{im}, \mu_m) = \frac{1}{2} \zeta_{im}^2 r_i^2(0) (1-\mu_m^2) \omega^2.
\end{equation}
Note that $Q$ is in unit of m$^2$ s$^{-2}$. Because $V$ is dimensionless, $Q$ should also be converted into dimensionless planetary unit as
\begin{equation}
    Q_{\rm pu} = \left(\frac{R_{\rm eq}}{GM}\right) Q_{\rm SI}.
\end{equation}
For a discussion on centrifugal potential assuming differential rotation on cylinders, see \cite{militzer_models_2019}. For now, \texttt{CORGI} assumes uniform rotation, while future implementations will incorporate differential rotation to explore the effects of deep atmospheric winds on the gravity field of Uranus.

\subsection{Analytical Expression for Newton Step} \label{sec:appendix_newton_step}
At each iteration, the CMS method updates the shape function $\zeta_{im}$ using a Newton step, which requires an analytical expression for the derivative $f'_{im}(\zeta_{im}) = df_{im}(\zeta_{im}) / d\zeta_{im}$. Recall that $f_{im}(\zeta_{im}) \equiv U_i(\zeta_{im}, \mu_m) - U_i(1, 0)$, where $U_i(1, 0)$ is independent of the shape function, so
\begin{equation}
    f'_{im}(\zeta_{im}) = \frac{d}{d\zeta_{im}} U_i(\zeta_{im},\mu_m) = \frac{d}{d\zeta_{im}} V_i(\zeta_{im},\mu_m) + \frac{d}{d\zeta_{im}} Q_i(\zeta_{im},\mu_m),
\end{equation}
where $V_i(\zeta_{im},\mu_m)$ is defined in Equation (\ref{eq:Vi_definition}) and $Q_i(\zeta_{im},\mu_m)$ is defined in Equation (\ref{eq:Qi_deifinition}). The derivative of the centrifugal potential is
\begin{equation} \label{eq:dQ_dzeta}
\begin{split}
    \frac{d}{d\zeta_{im}} Q_i(\zeta_{im},\mu_m) = \zeta_{im} r_i^2(0) (1-\mu_{m}^2) \omega^2.
\end{split}
\end{equation}
Differentiating Equation (\ref{eq:Vi_definition}) requires the chain rule. We first differentiate the prefactor
\begin{equation}
    \frac{d}{d\zeta_{im}} \left( -\frac{1}{\zeta_{im}\lambda_i} \right) = \frac{1}{\zeta_{im}^2 \lambda_i},
\end{equation}
and then rewrite $V_i$ in the following form for simplicity
\begin{equation}
    V_i(\zeta_i, \mu) = -\frac{1}{\zeta_i \lambda_i} 
    \left[
    \sum_{j=i}^{N_L-1} \sum_{n=0}^{\infty} S_1 + \sum_{j=0}^{i-1} \sum_{n=0}^\infty S_2 + \sum_{j=0}^{i-1} S_3
    \right].
\end{equation}
Now the sums can be differentiated individually as
\begin{equation}
\begin{split}
    \frac{d}{d\zeta_{im}} \sum_{j=i}^{N_L-1} \sum_{n=0}^{\infty} S_1 &= \frac{1}{\zeta_{im}} \sum_{j=i}^{N_L-1} \sum_{n=0}^{\infty} (-n) \widetilde{J}_{j,n} \left( \frac{\lambda_j}{\lambda_i \zeta_{im}} \right)^n P_n(\mu_m), \\ 
    \frac{d}{d\zeta_{im}} \sum_{j=0}^{i-1} \sum_{n=0}^\infty S_2 &= \frac{1}{\zeta_{im}} \sum_{j=0}^{i-1} \sum_{n=0}^\infty (n+1) \widetilde{J}_{j,n}' \left( \frac{\lambda_j \zeta_{im}}{\lambda_j} \right)^{n+1} P_n(\mu_m), \\ 
    \frac{d}{d\zeta_{im}} \sum_{j=0}^{i-1} S_3 &= \frac{1}{\zeta_{im}} \sum_{j=0}^{i-1} 3 \widetilde{J}_{j,0}'' \lambda_i^3 \zeta_{im}^3,
\end{split}
\end{equation}
where the summands remained unchanged except for extra factors involving $n$ and constants introduced by differentiation. Therefore,
\begin{equation}
\begin{split}
    \frac{d\Sigma S_1}{d\zeta_{im}} &\equiv \frac{d}{d\zeta_{im}} \sum_{j=i}^{N_L-1} \sum_{n=0}^{\infty} S_1 = \frac{1}{\zeta_{im}} \sum_{j=i}^{N_L-1} \sum_{n=0}^{\infty} (-n) S_1, \\ 
    \frac{d\Sigma S_2}{d\zeta_{im}} &\equiv \frac{d}{d\zeta_{im}} \sum_{j=0}^{i-1} \sum_{n=0}^\infty S_2 = \frac{1}{\zeta_{im}} \sum_{j=0}^{i-1} \sum_{n=0}^\infty (n+1) S_2, \\ 
    \frac{d\Sigma S_3}{d\zeta_{im}} &\equiv \frac{d}{d\zeta_{im}} \sum_{j=0}^{i-1} S_3 = \frac{1}{\zeta_{im}} \sum_{j=0}^{i-1} 3 S_3.
\end{split}
\end{equation}
With these simplifications, we may obtain an expression for $dV_i(\zeta_{im},\mu_m)/d\zeta_{im}$ using the chain rule as
\begin{equation} \label{eq:dV_dzeta}
\begin{split}
    \frac{d}{d\zeta_{im}} V_i(\zeta_{im},\mu_m) &= \frac{1}{\zeta_{im}^2 \lambda_i} \cdot (\Sigma S_1 + \Sigma S_2 + \Sigma S_3) -\frac{1}{\zeta_{im}\lambda_i} \left(\frac{d\Sigma S_1}{d\zeta_{im}} + \frac{d\Sigma S_2}{d\zeta_{im}} + \frac{d\Sigma S_3}{d\zeta_{im}}\right) \\
        &= \frac{1}{\zeta_{im}^2 \lambda_i} \left[
    \sum_{j=i}^{N_L-1} \sum_{n=0}^{\infty} (n+1) S_1 + \sum_{j=0}^{i-1} \sum_{n=0}^\infty (-n) S_2 + \sum_{j=0}^{i-1} (-2) S_3
    \right].
\end{split}
\end{equation}
An analytical expression for $f'_{im}(\zeta_{im})$ can now be obtained by combining Equation (\ref{eq:dQ_dzeta}) and (\ref{eq:dV_dzeta}).

\section{CMS Model Validation and Convergence} \label{sec:appendix_CMS_validation}
Our CMS model is validated by comparing to $J_n$ calculated for a $n=1$ polytrope Jupiter model presented in \cite{wisdom_differential_2016}. The polytropic Jupiter model presented in \cite{wisdom_differential_2016} was computed using CMS and two other independent methods: the CLC method and a method applying spherical Bessel functions. The Bessel function method was solvable thanks to the $n=1$ polytrope assumption. The Bessel function solutions have high precision to at least 16 digits and are therefore essentially exact, providing a reliable ground for comparison. 

\begin{table*}
    \centering
    \begin{tabular}{llll}
    \hline
         &  CMS ($N_L=128$, $N_M$=48) & Bessel function & Relative error \\
    \hline
         $J_2$    & $1.400222252437312 \times 10^{-2}$ & $1.398851089834702 \times 10^{-2}$ & $9.802062653943783 \times 10^{-4}$ \\
         $J_4$    & $-5.327791129361269 \times 10^{-4}$ & $-5.318281001092907 \times 10^{-4}$ & $1.788195897585813 \times 10^{-3}$ \\
         $J_6$    & $3.019606862436119 \times 10^{-5}$ & $3.011832290533641 \times 10^{-5}$ & $2.581342901101601 \times 10^{-3}$ \\
         $J_8$    & $-2.139307027954382 \times 10^{-6}$ & $-2.13211571072505 \times 10^{-6}$ & $3.372855044009958 \times 10^{-3}$ \\
         $J_{10}$ & $1.747922200581373 \times 10^{-7}$ & $1.740671195866297 \times 10^{-7}$ & $4.165637216434328 \times 10^{-3}$ \\
         $J_{12}$ & $-1.575998140569955 \times 10^{-8}$ & $-1.568219505562893 \times 10^{-8}$ & $4.960169784567541 \times 10^{-3}$ \\
         $J_{14}$ & $1.526837910969635 \times 10^{-9}$ & $1.518099226841379 \times 10^{-9}$ & $5.756332638702814 \times 10^{-3}$ \\
         $J_{16}$ & $-1.562156435854803 \times 10^{-10}$ & $-1.551985081630105 \times 10^{-10}$ & $6.553770616155035 \times 10^{-3}$ \\
         $J_{18}$ & $1.668101162035406 \times 10^{-11}$ & $1.655925984019243 \times 10^{-11}$ & $7.352489261996578 \times 10^{-3}$ \\
         $J_{20}$ & $-1.843476916630424 \times 10^{-12}$ & $-1.828574676494702 \times 10^{-12}$ & $8.149648098752566 \times 10^{-3}$ \\
    \hline
    \end{tabular}
    \caption{CMS results for Jupiter compared to Bessel function solutions from \cite{wisdom_differential_2016}. The Bessel function solutions are precise to at least 16 digits and essentially provide ground truth for comparison. Small relative error validates our CMS model.}
    \label{tab:CMS_vs_Bessel}
\end{table*}

\begin{figure}
    \centering
    \includegraphics[width=0.6\linewidth]{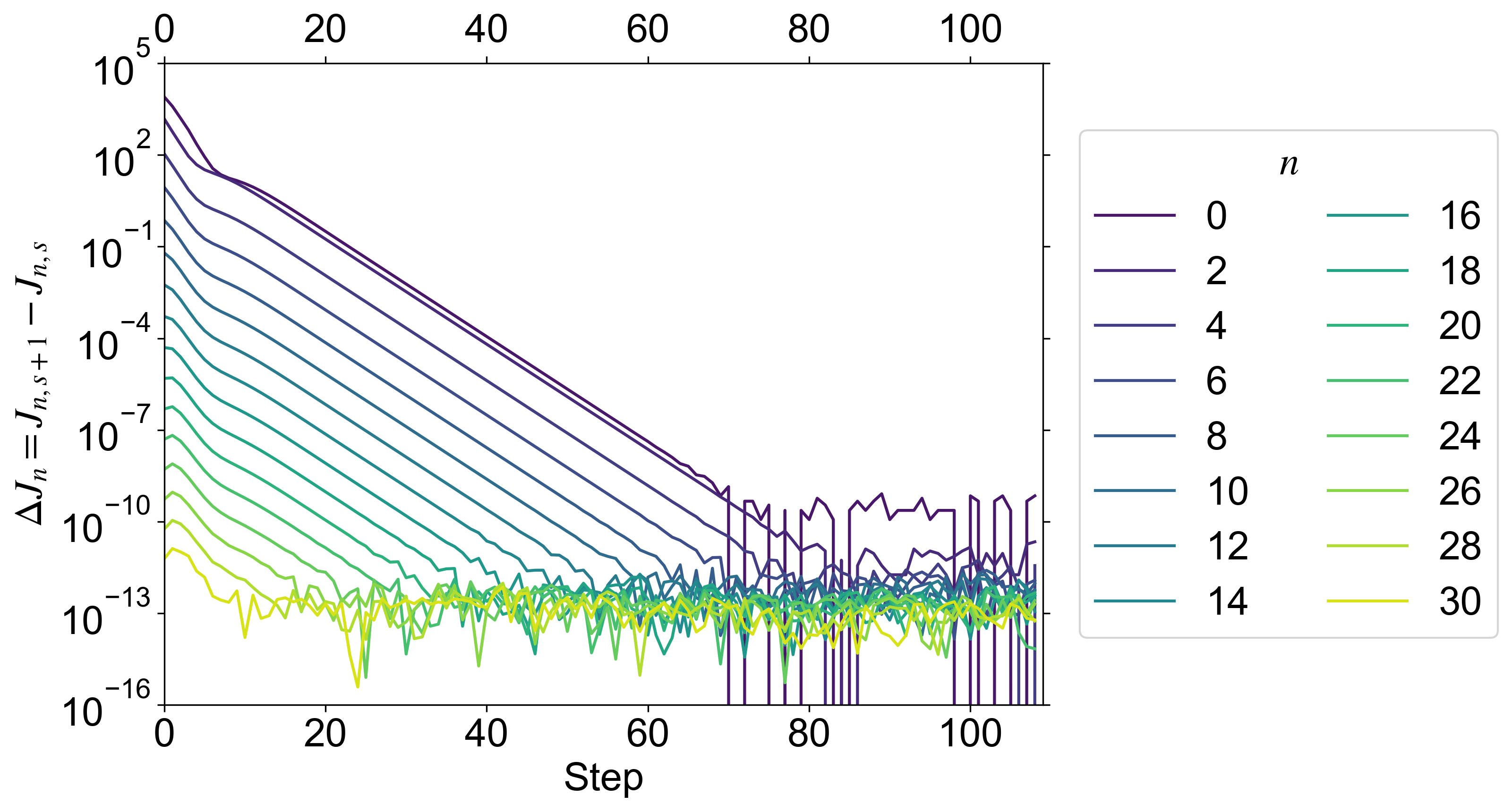}
    \caption{CMS model convergence for the validation run (Jupiter model with $n=1$ polytrope, $N_L=128$, $N_M=48$). The lines show the change in $J_n$ between consecutive steps, defined as $\Delta J_n = J_{n,s+1} - J_{n,s}$, where $s$ is the step number. Colors represent (from darkest to lightest) $n=0$ to $n=30$ harmonics. The higher order harmonics reach convergence with $\Delta J_n \sim 10^{-13}$ more rapidly, while $J_2$ is converged after $\sim80$ steps with $\Delta J_n \sim 10^{-11}$, comparable to the theoretical precision limit of Gaussian quadrature ($\sim10^{-12}$).}
    \label{fig:jupiter_jn_convergence}
\end{figure}

Basic input parameters for the Jupiter model are as follows. The rotation period is 9$^\text{h}$55$^\text{m}$29.7$^\text{s}$, $R_{\rm eq} =$ 71,492 km, and $GM = 126686536.1$ km$^3$ s$^{-2}$. The polytropic EOS is given by
\begin{equation}
    P = K\rho^{1 + \frac{1}{n}},
\end{equation}
with $n=1$. The constant $K$ is found iteratively using a bisection method, such that the $\rho(P)$ profile integrates to the correct total mass. Gravity harmonics $J_n$ computed by our CMS model is shown in Table \ref{tab:CMS_vs_Bessel}, along with Bessel function solutions from \cite{wisdom_differential_2016} and relative errors. The relative error of $J_2$ is approximately $9.80 \times 10^{-4}$, while the relative errors for all other $J_n$ harmonics are on the order of $10^{-3}$. For comparison, the $J_2\times10^6$ and $J_4\times10^6$ measured by V2 for Uranus are $3510.7\pm0.7$ ($1.99\times10^{-4}$ relative error) and $-34.2\pm1.3$ ($3.80\times10^{-2}$ relative error), respectively \citep{jacobson_orbits_2014}. Therefore, our CMS model have comparable or better precision than available observation and is sufficient for simulating the gravity harmonics of hypothetical Uranus interior models to inform UOP gravity field measurements. CMS precision can be easily improved by increasing the number of layers, as shown by the CMS runs with $N_L=512$ layers in \cite{wisdom_differential_2016}, which have $\sim10$ times less relative error than our $N_L=128$ model. However, CMS computation time increases rapidly with $N_L$. To perform rapid parameter space sweep for hundreds of forward models, we stick to $N_L=128$. The choice of $N_L=128$ is also justified by the goal of our study. We do not intend to find Uranus interior models that exactly fit the V2 measurements, which would require much smaller CMS uncertainty than measurement uncertainty. Rather, we aim to explore the $J_n$ parameter spaces covered by different compositional categories. Due to the intrinsic compositional degeneracy within each category, simulated $J_n$ data points scatter around a parameter space much larger than measurement uncertainty or CMS uncertainty (Figure \ref{fig:j2j4_error_bar} and \ref{fig:j6j8_error_bar}). In addition, interpretation of V2 data is itself uncertain. Recent reanalysis by \cite{french_uranus_2024} differ from previous analysis by \cite{jacobson_orbits_2014}. The difference is significant in $J_4$ and dwarfs CMS uncertainty (Figure \ref{fig:j2j4_error_bar}b and c). For our purpose, therefore, uncertainty introduced by the $N_L=128$ model is insignificant.

Convergence of our CMS model is shown in Figure \ref{fig:jupiter_jn_convergence}. All $\Delta J_n$ decreases logarithmically with step until reaching convergence, after which they oscillate around the numerical precision limit. Higher order $J_n$ reach convergence more rapidly. $J_2$ is converged after $\sim80$ steps, reaching a precision of $\sim10^{-11}$, close to the theoretical precision limit of Gaussian quadrature ($\sim10^{-12}$). Running a CMS model with $N_L=128$ and $N_M=48$ until convergence requires $\sim67$ hours ($\sim3000$ s per step) on a Intel Xeon Platinum 8260 processor with 2.40 GHz base frequency.

\bibliography{uranus_interior}{}
\bibliographystyle{aasjournal}



\end{document}